\documentclass[10pt,sigplan]{acmart}
\usepackage[10pt]{sigmin}
\usepackage{tikz}
\usepackage{circledsteps}

\usepackage{amsmath}
\usepackage[ruled,vlined]{algorithm2e}
\usepackage{caption}
\usepackage{graphicx}
\usepackage{subcaption}
\usepackage{circledsteps}
\usepackage{booktabs}
\usepackage{hyperref}%
\usepackage{pifont}
\usepackage[english]{babel}
\usepackage{makecell}
\usepackage{color}
\usepackage[misc]{ifsym}
\usepackage{bbding}
\usepackage{breakurl}

\makeatletter
\renewcommand{\sectionautorefname}{\S\@gobble}
\renewcommand{\subsectionautorefname}{\S\@gobble}
\renewcommand{\subsubsectionautorefname}{\S\@gobble}
\makeatother

\hypersetup{
  colorlinks=true,
  linkcolor=blue!90!red,
  citecolor={red!60!black},
  urlcolor={blue!60!black}
}

\usepackage[labelformat=simple]{subcaption}

\usepackage{booktabs}
\usepackage{enumitem}

\usepackage{multirow}

\setlist{topsep=0pt, itemsep=0pt, parsep=0pt, leftmargin=1em}

\newcommand{\sys}{SwitchFS}

\newsavebox{\twosubbox}

\copyrightyear{2026}
\acmYear{2026}
\setcopyright{cc}
\setcctype{by}
\acmConference[EUROSYS '26]{21st European Conference on Computer Systems}{April 27--30, 2026}{Edinburgh, Scotland Uk}
\acmBooktitle{21st European Conference on Computer Systems (EUROSYS '26), April 27--30, 2026, Edinburgh, Scotland Uk}
\acmPrice{}
\acmDOI{10.1145/3767295.3769349}
\acmISBN{979-8-4007-2212-7/26/04}

\begin{document}


\title{{\sys}: Asynchronous Metadata Updates for Distributed Filesystems with In-Network Coordination}

\author{\rm Jingwei Xu,\; Mingkai Dong,\; Qiulin Tian,\; Ziyi Tian,\; Tong Xin,\; and Haibo Chen\\
  Institute of Parallel and Distributed Systems, Shanghai Jiao Tong University
} %

\begin{abstract}
Distributed filesystem metadata updates are typically synchronous.
This creates inherent challenges for access efficiency, 
load balancing, and directory contention, especially under dynamic and skewed workloads.
This paper argues that synchronous updates are overly conservative.
We propose {\sys} with \emph{asynchronous metadata updates} that allow operations to return early and defer directory updates until reads, both hiding latency and amortizing overhead.
The key challenge lies in efficiently maintaining the synchronous POSIX semantics of metadata updates. 
To address this, {\sys} is co-designed with a programmable switch, leveraging the limited on-switch resources to track directory states with negligible overhead.
This allows {\sys} to aggregate and apply delayed updates efficiently, using batching and consolidation before directory reads.
Evaluation shows that {\sys} achieves up to 13.34$\times$ and 3.85$\times$ higher throughput, and 61.6\% and 57.3\% lower latency than two state-of-the-art distributed filesystems, Emulated-InfiniFS and Emulated-CFS, respectively, under skewed workloads.
For real-world workloads, {\sys} improves end-to-end throughput by 21.1$\times$, 1.1$\times$, and 0.3$\times$ over CephFS, Emulated-InfiniFS, and Emulated-CFS, respectively.

\end{abstract}



\maketitle
\pagestyle{plain}
\section{Introduction}%
\label{sec:intro}

Metadata performance is a critical factor in the efficiency of distributed filesystems (DFSs) deployed in modern data centers.
To manage the filesystem namespace and file attributes, modern DFSs~\cite{Weil2006Ceph,BeeGFS,Shvachko2010The-Hadoop,Lv2022InfiniFS,Li2017LocoFS,Pan2021Facebooktextquoterights,CFS,Ren2014IndexFS} typically employ a dedicated metadata service.
Before accessing file data, clients must first contact this service to check permissions and obtain the data location.
Because files in data centers are often small~\cite{CFS,10.1109/IPDPS.2009.5161029,Beaver2010Finding,Li2017LocoFS,Wang2021Lunule,10.5555/2591305.2591325,10.5555/1267724.1267728,10.1145/3366684,10.1109/TPDS.2020.3034517}, the relative overhead of metadata operations becomes significant.
For instance, traces from Baidu AI Cloud~\cite{CFS} report that metadata operations account for 67\%--96\% of all filesystem requests and dominate I/O completion time.
Unlike data access bandwidth, which scales nearly linearly with the number of data servers, metadata throughput is difficult to scale due to hierarchical dependencies~\cite{CFS,Lv2022InfiniFS,Ren2014IndexFS,Li2017LocoFS}.
As a result, metadata frequently becomes the performance bottleneck.
In large-scale applications such as deep learning model training~\cite{FalconFS-arxiv,Mantle} and data analytics~\cite{Mantle}, this bottleneck can lead to underutilized data bandwidth and compute resources, ultimately degrading end-to-end performance~\cite{FalconFS-arxiv,Mantle,Wang2021Lunule}.

A series of studies~\cite{Weil2006Ceph,TableFS,Ren2014IndexFS,Li2017LocoFS,Niazi2017HopsFS,Lv2022InfiniFS,Pan2021Facebooktextquoterights,CFS,Xfast,MetaWBC,DelveFS,hashFS} have been made on scaling DFS metadata performance, yet there are still challenges remaining.
First, while DFSs typically partition the directory tree across multiple metadata servers to scale out, selecting partitioning granularity presents an inherent trade-off between load balance and operation overhead, making it hard to achieve both~\cite{Macko2022Survey}.
Coarse-grained partitioning~\cite{Weil2006Ceph,Lv2022InfiniFS,Niazi2017HopsFS,Pan2021Facebooktextquoterights,BeeGFS,Ren2014IndexFS} groups file inodes with their parent directory, making large directories easily become hotspots, especially in skewed workloads~\cite{Wang2021Lunule,Abad2012Metadata}.
In contrast, fine-grained partitioning~\cite{Gluster2019,Li2017LocoFS,CFS} evenly shuffles file inodes across metadata servers, achieving good load balance but breaking the colocation between inodes and parent directories;
as metadata is partitioned across different servers, metadata operations need to use (costly) distributed transactions to update them consistently.
Second, state-of-the-art distributed metadata services often suffer from contention on directory metadata.
Specifically, parent-related operations (e.g., \emph{mkdir} and \emph{create}) contend on the parent directory, resulting in degraded scalability~\cite{CFS,SingularFS}.

A fundamental reason for the aforementioned challenges is synchronous metadata updates.
As filesystem semantics require durable visibility (the effect of an operation is visible to subsequent operations) and atomicity (no partial updates), state-of-the-art DFSs typically use synchronous transactions.
Unfortunately, this approach incurs cross-server coordination overhead and metadata contention on the critical path, leading to high latency and throughput bottlenecks.

In this paper, we argue that \emph{synchronous updates are overly conservative}.
We observe that \emph{directories are typically not read immediately after being updated}.
It provides an opportunity for \emph{asynchronous updates}, deferring directory updates until the next read, to hide update latency and to batch updates.

However, to preserve the durable visibility of metadata updates, a mechanism is required to identify directories with pending updates to avoid stale reads.
A straightforward approach would be to dedicate a server on the critical path of metadata operations to track directory updates and to notify directory reads.
It introduces an additional RTT, and the dedicated server can become a bottleneck.
In contrast, we find that the programmable switch --- naturally positioned on the critical path of metadata operations and capable of extremely high packet processing throughput --- is an ideal candidate to coordinate asynchronous updates and reads.

Based on these insights, we present \textbf{{\sys}}, an asynchronous distributed filesystem co-designed with a program\-mable switch for operation coordination.
{\sys} is POSIX-compliant, with the goal of achieving good load balance, low latency, and high scalability under skewed workloads at the same time.
{\sys} has the following design features.

First, we propose an efficient protocol for asynchronous metadata updates%
, offering two benefits:
(a) it enables most metadata operations to complete in a single RTT while using fine-grained partitioning for load balance, resolving the trade-off faced by previous DFSs;
(b) it merges contending metadata updates before applying them, mitigating metadata contention.
Specifically, when an operation updates objects across multiple servers (typically a file and its parent directory),
{\sys} logs the directory update on the server hosting the file's inode and marks the directory as \emph{scattered} in the programmable switch.
When reading a directory, the switch signals if the directory is scattered.
If so, {\sys} \emph{aggregates} the deferred updates from other servers before accessing the directory.
To further expedite the aggregation process and reduce contention on directory metadata,
{\sys} consolidates metadata updates to the same directory before aggregation, leveraging the inherent commutativity of updates to directory attributes~\cite{180269,CFS,10.1145/3068914,10.1145/2757667.2757683,10.1145/3447865.3457970,10.1145/3465332.3470872,ahmednacer2012filecrdt}.

Second, we leverage a programmable switch to efficiently track scattered directories within the network.
Our insight is that, thanks to its high packet processing capability and central position in the network, the programmable switch serves as a superior coordinator compared to a standard server.
Given the limited on-switch memory and compute resources, we design a compact in-network dirty set with high memory utilization and low conflict rate by organizing the switch registers in a set-associative manner.
By inspecting specific headers in packets, the dirty set can insert, query, and remove directory \emph{fingerprints} at line rate, and redirect packets as needed, enabling {\sys} to query and update directory states at negligible overhead.

\begin{figure}[tb]
  \centering
  \begin{subfigure}[b]{0.49\linewidth}
      \centering
      \includegraphics[width=0.7\linewidth]{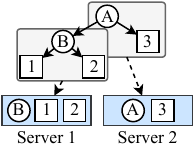}
      \vspace{-7pt}
      \caption{P/C grouping}
      \label{fig:partition-1}
  \end{subfigure}
  \begin{subfigure}[b]{0.49\linewidth}
      \centering
      \includegraphics[width=0.7\linewidth]{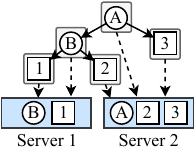}
      \vspace{-7pt}
      \caption{P/C separation}
      \label{fig:partition-2}
  \end{subfigure}
\caption{\textbf{Metadata partitioning approaches.}}
\label{fig:partition}
\end{figure}

Our evaluation shows that {\sys} achieves up to 13.34$\times$ higher throughput than Emulated-InfiniFS~\cite{Lv2022InfiniFS}, and 57.3\% lower latency than Emulated-CFS~\cite{CFS}. %
For real-world workloads, {\sys} improves end-to-end throughput by 21.1$\times$, 1.1$\times$, and 0.3$\times$ over CephFS, Emulated-InfiniFS, and Emulated-CFS, respectively.

In summary, this paper makes the following contributions.

\begin{itemize}[itemsep=0pt,leftmargin=1em]
\item {\bf An analysis of key challenges in scaling DFS metadata performance (\autoref{sec:Challenges}).}
We identify that the inherent trade-off in namespace partition granularity and the directory contention makes metadata performance suboptimal.
\item {\bf A distributed filesystem, {\sys}, adopting asynchronous metadata updates (\autoref{sec:design}).}
{\sys} resolves the trade-off between operation efficiency and load balancing without compromising consistency with asynchronous metadata updates (\autoref{sec:design-asynchronous-metadata-operations}), and mitigates contention over directory metadata with change-log compaction (\autoref{sec:design-aggregation}).

\item {\bf In-network dirty set that tracks directory states (\autoref{sec:design_of_switch_data_plane}).}
To the best of our knowledge, {\sys} is the first to incorporate in-network optimization into a DFS.
\end{itemize}

\section{Background}

\subsection{Distributed Metadata Partitioning}
\label{sec:Existing Distributed Metadata Services}

\begin{table}[tb]
  \centering
  \resizebox{\linewidth}{!}{%
  \begin{tabular}{ccc}
  \toprule
  & P/C Grouping & P/C Separation  \\
  \midrule
  Single-inode Op.   & Local                 & Local        \\
  Double-inode Op.  & Local/Cross-server  & Cross-server    \\
  Load Balancing    & Directory hotspots    & Balanced   \\
  Example FS       & CephFS~\cite{Weil2006Ceph}, InfiniFS~\cite{Lv2022InfiniFS} & GlusterFS~\cite{Gluster2019}, CFS~\cite{CFS}   \\
  \bottomrule
  \end{tabular}
  }
  \caption{\textbf{Trade-offs among partitioning approaches. "P/C" stands for "parent-children".} With P/C grouping, \emph{mkdir}, \emph{rmdir} are cross-server. With P/C separation, \emph{mkdir}, \emph{rmdir}, \emph{create}, \emph{delete} are cross-server.}
  \label{tab:partition}
\end{table}

Many DFSs partition the directory tree across a cluster of metadata servers to scale out metadata service~\cite{Weil2006Ceph,Lv2022InfiniFS,Niazi2017HopsFS,CFS,Pan2021Facebooktextquoterights,Li2017LocoFS}.
The partitioning approach is an important design choice that impacts operation overhead and load distribution, and ultimately, overall performance and scalability.
Existing approaches can be classified into \emph{parent-children grouping} (\emph{P/C grouping}) and \emph{parent-children separation} (\emph{P/C separation}), depending on whether file inodes are colocated with their parent directories.
The approaches are illustrated in \autoref{fig:partition} and compared in \autoref{tab:partition}.

\paragraph{Parent-children grouping.}
\emph{P/C grouping} colocates file inodes with their parent directories, while directory inodes may be distributed independently for scalability, e.g., subtree partitioning~\cite{Ghemawat2003The-Google,Shvachko2010The-Hadoop} and per-directory hashing~\cite{BeeGFS,Pan2021Facebooktextquoterights,Lv2022InfiniFS,Ren2014IndexFS,Niazi2017HopsFS}, as illustrated in \autoref{fig:partition-1}.
These studies adopt P/C grouping since it achieves good metadata locality, thus reducing operation overhead.
As metadata operations often simultaneously update metadata object and its parent directory, e.g., file \emph{create},
with parent and children colocated, some updates are performed on the same server, avoiding costly cross-server coordination.
The downside of P/C grouping is that it places all file inodes within the same directory onto a single server, leading to substantial load imbalance in skewed workloads, thereby undermining scalability.

\paragraph{Parent-children separation.}
\emph{P/C separation} distributes file inodes independently of the location of their parent directories, e.g., per-file hashing~\cite{Gluster2019,CFS}, as illustrated in \autoref{fig:partition-2}.
By evenly distributing metadata objects, \emph{P/C separation} achieves good load balance.
However, metadata operations suffer from expensive cross-server coordination to update multiple objects across servers consistently.

\subsection{Programmable Switch}
A programmable switch is a packet-switching device equip\-ped with programmable ASICs and stateful memory~\cite{Tofino,XPliant,Cisco}, enabling in-network co-design for distributed systems.
It can do user-defined manipulation on packets while forwarding them at extremely high speeds.
Compared with manipulating packets with a standard server, the program\-mable switch has two advantages:
(a) its central location in the network grants it a global view of all the communications;
(b) the ASIC-based hardware processes packets at line rate, providing high throughput and low latency.
We support these statements with our evaluation in \autoref{eval:Contribution pswitch}.

\section{Motivation and Challenges}

\subsection{Characteristics of Datacenter Workloads}
\label{sec:Characteristics of Data Center Workloads}

\begin{table}[t]
  \centering
  \footnotesize
  \begin{tabular}{ll|llll} 
  \toprule
  Category                                                                   & \multicolumn{1}{c}{Ratio} & \multicolumn{4}{c}{Detailed Operation Ratio}   \\ 
  \hline
  \multirow{3}{*}{\begin{tabular}[c]{@{}l@{}}Dir. Update\end{tabular}} & \multirow{3}{*}{30.76\%}  & create      & 9.58\% & delete  & 11.88\%  \\
                                                                             &                           & mkdir       & 0.01\%  & rmdir   & 0.01\%   \\
                                                                             &                           & file rename & 9.29\% &         &          \\ 
  \hline
  \begin{tabular}[c]{@{}l@{}}Dir. Read\end{tabular}                    & 4.19\%                    & statdir     & 0.28\%  & readdir & 3.91\%  \\ 
  \hline
  \multirow{2}{*}{Others}      & \multirow{2}{*}{65.05\%}  & open/close  & 52.59\% & stat    & 12.35\%  \\ 
                               &                           & others       & 0.10\%  &         &          \\
  \hline
  \end{tabular}
  \caption{\textbf{Ratio of metadata operations in workloads from deployed PanguFS instances in Alibaba~\cite{Lv2022InfiniFS}.}}
  \label{tab:operation-ratio}
\end{table}

\paragraph{A directory is typically not read immediately after being updated.}
\autoref{tab:operation-ratio} presents the ratio of metadata operations in three deployed PanguFS instances in Alibaba~\cite{Lv2022InfiniFS} under various workloads like data processing, object storage, and block storage.
Operations that update directories (i.e., \emph{create}, \emph{delete}, \emph{mkdir}, \emph{rmdir}, and \emph{rename}) constitute 30.76\% of all metadata operations, while operations that read directories (i.e., \emph{readdir} and \emph{statidr}) constitute merely 4.19\%.
By the pigeonhole principle, the disparity implies that at least $(30.76\% - 4.19\%) / 30.76\% = 86.3\%$ of the directory updates are not immediately followed by a directory read, satisfying that either
(a) the updated directory is never read after the update or
(b) the updated directory is updated by at least one other operation before it is read.
Thus, applying directory updates in a delayed manner can help hide latency and mitigate contention.
Notably, the partial order of dependent updates (e.g., creation and deletion of the same file) should be preserved, which {\sys} guarantees.

\paragraph{Datacenter workload is skewed along multiple dimensions.}
This skewness challenges metadata management and can hinder performance scaling if the load balance is poor.

\begin{itemize}[itemsep=0pt,leftmargin=1em]
\item \emph{Unbalanced directory hierarchies.}
Analyses of deployed DFSs~\cite{Ren2014IndexFS,Dayal2008CharacterizingHS,10.14778/2536206.2536213} show that while most directories contain fewer than 128 entries, some directories include over a million entries.
Such an unbalanced hierarchy complicates the balancing of inode distribution and metadata load.

\item \emph{Directory hotspots.}
In addition,
applications commonly group related files into directories and access them within short temporal intervals~\cite{Pan2021Facebooktextquoterights,Li2017LocoFS,Lv2022InfiniFS}.
This creates temporal hotpots, which further complicates load balancing.

\end{itemize}

\subsection{Challenges for Metadata Service}
\label{sec:Challenges}

\begin{figure}[t]
  \centering
  \begin{subfigure}[b]{0.49\linewidth}
      \centering
      \includegraphics[width=\linewidth]{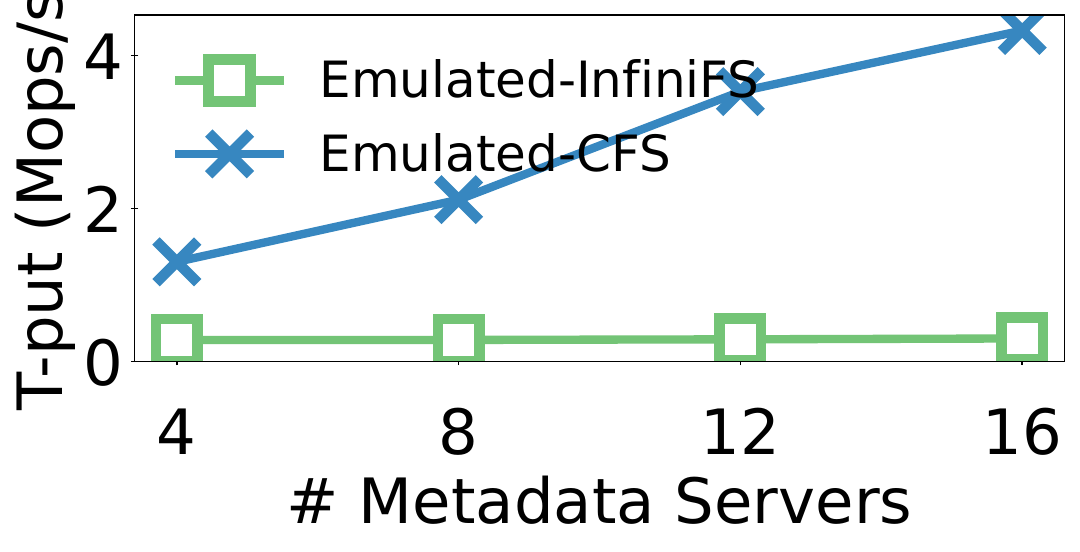}
      \vspace{-15pt}
      \caption{Throughput of \emph{stat}.}
      \vspace{10pt}
      \label{fig:meta stat throughput}
  \end{subfigure}
  \begin{subfigure}[b]{0.49\linewidth}
    \centering
    \includegraphics[width=\linewidth]{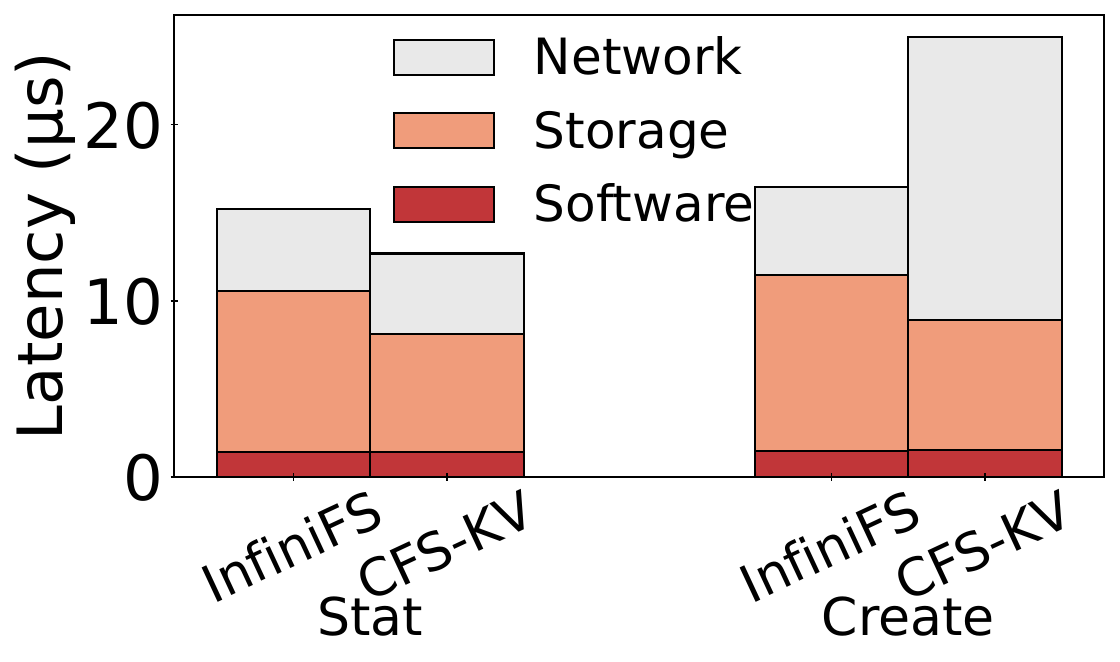}
    \vspace{-15pt}
    \caption{Latency breakdown.}
    \vspace{10pt}
    \label{fig:meta latency breakdown}
  \end{subfigure}
  \begin{subfigure}[b]{0.49\linewidth}
      \centering
      \includegraphics[width=\linewidth]{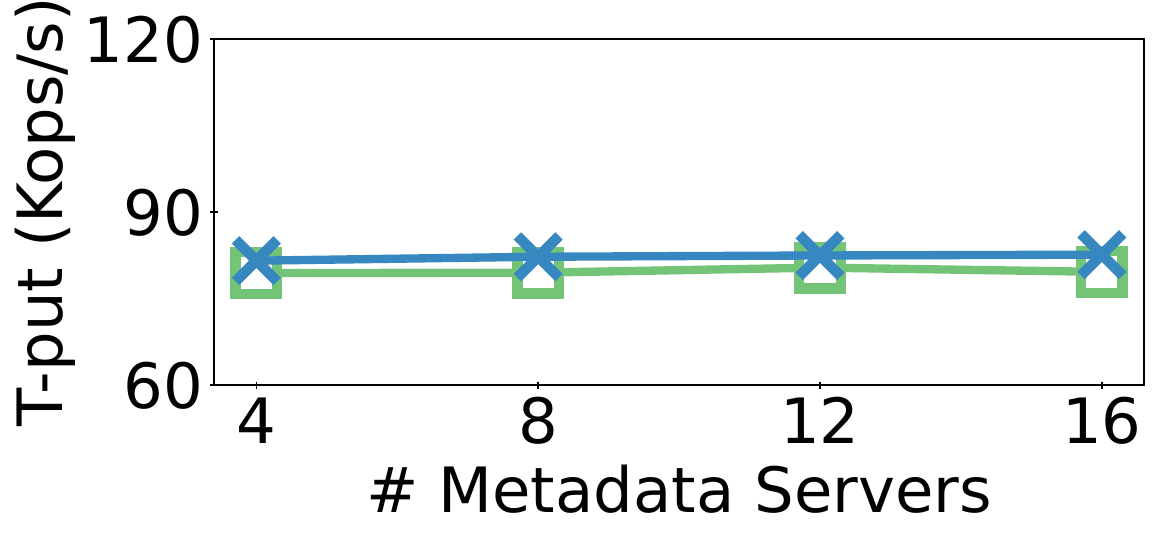}
      \vspace{-15pt}
      \caption{Throughput of \emph{create} with varying numbers of servers.}
      \label{fig:meta create throughput}
  \end{subfigure}
  \begin{subfigure}[b]{0.49\linewidth}
    \centering
    \includegraphics[width=\linewidth]{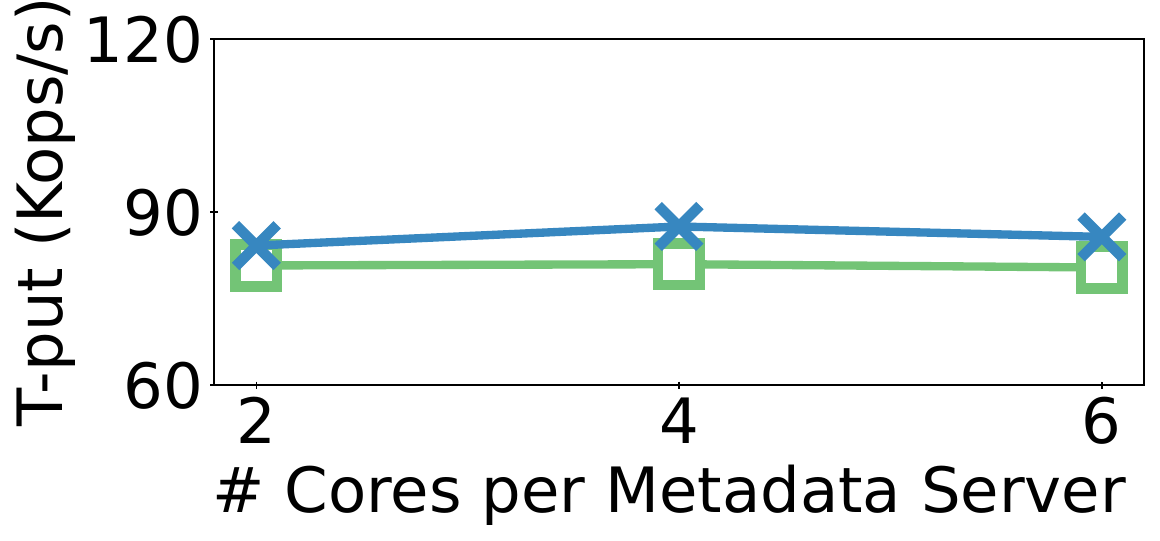}
    \vspace{-15pt}
    \caption{Throughput of \emph{create} with varying \# of cores per server.}
    \label{fig:meta-create-throughput-core}
  \end{subfigure}
\caption{\textbf{Throughput scalability and latency of metadata operations.} Operations are performed in a shared directory. DFSs have eight servers, each with four cores by default.
}
\label{fig:meta analyses}
\end{figure}

Modern data center applications demand DFS metadata operations with high scalability, high throughput, and low latency.
However, there are several challenges in the design.

\paragraph{\underline{Challenge \#1:}} \emph{There is an inherent trade-off between low operation overhead and load balance in the design of directory tree partitioning.}

To reduce operation overhead, it is important to colocate files' inodes with the parent directory.
Metadata operations (e.g., \emph{create} and \emph{delete}) often update a metadata object together with its parent directory's metadata.
With the objects colocated, the operation can be executed on a single server, avoiding expensive cross-server coordination~\cite{Lv2022InfiniFS}.

However, as files in the same directory are often related and accessed in a short time, grouping them makes the filesystem vulnerable to hotspots~\cite{Pan2021Facebooktextquoterights,Li2017LocoFS,Lv2022InfiniFS}.
Therefore, to improve load balance, it is important to scatter files of the same directory across servers.

Unfortunately, colocating inherently conflicts with scattering, making it difficult to have the best of both worlds.

To evaluate how this trade-off affects performance, we compare the performance of Emulated-InfiniFS (E-InfiniFS)~\cite{Lv2022InfiniFS} and Emulated-CFS (E-CFS)~\cite{CFS}, two state-of-the-art DFSs adopting P/C grouping and P/C separation, respectively.
In \autoref{fig:meta stat throughput}, when performing \emph{stat} on uniformly random files in a shared directory, E-CFS's throughput scales linearly since files are evenly distributed across servers, while E-InfiniFS fails to scale because all file inodes reside on the same server, indicating that E-CFS has better load balance.
In contrast, in \autoref{fig:meta latency breakdown}, E-CFS suffers higher \emph{create} latency than E-InfiniFS, mainly due to higher network overhead from cross-server coordination, indicating that E-InfiniFS has lower operation overhead.

\paragraph{\underline{Challenge \#2:}}\emph{Concurrent operations in the same directory suffer low inter- and intra-server parallelism due to contention on the parent directory, hindering performance scaling.}

Regardless of the partitioning approach, the contention on directory metadata often becomes a performance bottleneck.
For example, when concurrently creating files in a single directory (e.g., during data preparation~\cite{FalconFS-arxiv}, compiling, decompression, etc.), each \emph{create} operation updates the attributes and entry list of the parent directory.
Although the creation of individual files could be parallelized across multiple servers (under P/C separation), updates to the parent directory are serialized at the server that stores it, resulting in low \emph{inter-server parallelism}.
Also, as conventional approaches rely on transactions and locks to guarantee the atomicity of metadata update, the server cannot leverage multicore parallelism, resulting in low \emph{intra-server parallelism}.

In \autoref{fig:meta create throughput} and \autoref{fig:meta-create-throughput-core}, both DFSs' throughput does not scale with the number of servers and of cores per server.
It indicates that both inter-server and intra-server parallelism are restricted by the contention on the parent directory.

\section{Overview}%
\label{sec:overview}
To overcome the inherent limitations in \autoref{sec:Challenges}, {\sys} coordinates asynchronous metadata operations in the network to promote parallelism and to hide latency.
We overview the design rationale and architecture in this section, then present the workflow in \autoref{sec:design} and the switch data plane in \autoref{sec:design_of_switch_data_plane}.

\subsection{Design Rationale}
\label{sec:Design Rationale}

\paragraph{Key idea: asynchronous directory updates.}
Our key idea is to \emph{leverage asynchronous updates to hide cross-server coordination overhead and exploit metadata operation parallelism}.
File-system operations (except for \emph{rename}, discussed in \autoref{sec:design-asynchronous-metadata-operations}) update at most two metadata objects: the target object and its parent directory~\cite{SingularFS};
thus, by deferring updates to the remote parent directory, {\sys} can execute operations locally on a single server.
This approach hides cross-server overhead and enables concurrent updates to a parent directory to execute in parallel.

\paragraph{Performance enabler: programmable switch.}
To ensure that a delayed update is visible to subsequent accesses, {\sys} leverages a centralized programmable switch to store directory state at line rate speed and sub-RTT latency.

However, due to limited on-chip memory and computational resources, it is non-trivial to maintain all directories' states for an entire filesystem.
We observe that, \emph{although the total number of directories can be huge, the set of directories whose metadata has been modified in a recent time interval is small}.
For example, consider a filesystem capable of creating or deleting one million files per second. 
The number of directories modified per 100 milliseconds is 100 thousand at most, small enough to fit into the switch's on-chip memory.
Therefore, by periodically applying delayed updates 
and clearing the on-switch data structure, on-switch state tracking is feasible.

\subsection{Overall Architecture}

As depicted in \autoref{fig:design-workflow-1}, {\sys} consists of three kinds of components: clients, metadata servers, and programmable switches.
Without loss of generality, we assume a single programmable switch for clarity, and defer the discussion of scaling to multiple switches to \autoref{sec:dataplane-deployment}.

\paragraph{Clients.}
{\sys} provides a user-space library (LibFS) for clients to communicate with servers.
The client maintains a metadata cache to accelerate path resolution, which caches only directory metadata.

\paragraph{Metadata servers.}
{\sys} distributes metadata objects across metadata servers with fine-grained partitioning.
A server stores its metadata in a \emph{key-value store} (i.e., Rocks\-DB\cite{2021RocksDB}).
The server tracks deferred modifications to remote directories in a per-directory \emph{change-log}, and tracks recently deleted directories in its \emph{invalidation list}.
The server uses a \emph{write-ahead log} (WAL) for crash recovery.

\paragraph{In-network dirty set.}
The programmable switch monitors network traffic to maintain a centralized \emph{dirty set} of delayed updates.
To efficiently utilize limited on-switch memory, the dirty set is stored in a multi-slot hash table structure.

\begin{table}[tb]
  \centering
  \resizebox{\linewidth}{!}{%
  \begin{tabular}{lccc}
  \toprule
   & Key                       & Value  & Partition By                                    \\
  \midrule
  Dir Metadata  & \emph{pid, name}          & \emph{id, timestamps, perm., etc.}  & the key \\
  Dir Entry  & \emph{pid, name}          & \emph{file type, permissions}  & N/A \\
  File Metadata       & \emph{pid, name}    & \emph{full regular file metadata} & the key             \\
  \bottomrule
  \end{tabular}
  }
  \caption{\textbf{Metadata schema.}
  Each directory has a 256-bit \emph{id}.
  \emph{pid} refers to the \emph{id} of the parent directory.
  A directory's entry list consists of multiple \emph{Dir Entry} key-value pairs, which are stored on the same server as the directory.
  \label{tab:metadata schema}
  }%
\end{table}

\subsection{Metadata Schema}
\label{sec:metadata schema}
\autoref{tab:metadata schema} presents the metadata schema of {\sys}.
Each directory is assigned a unique 256-bit identifier \emph{id} upon creation.
Metadata objects (i.e., inodes and dentries) are stored as key-value pairs, where the key is the concatenation of the parent directory's \emph{id} (\emph{pid}) and the file/directory \emph{name}, and the value is the metadata attributes.

{\sys} employs the P/C separation approach for load balance, partitioning file/directory metadata by hashing the \emph{(pid, name)} pair.
A directory's entry list is stored as multiple \emph{Dir Entry} key-value pairs, all of which are stored on the same server as the directory's inode.

Though not shown in the table, {\sys} generates a 49-bit \emph{fingerprint} for each directory by hashing \emph{(pid, directory name)}.
The fingerprint, which serves as the directory identifier within the switch, is designed to fit within the switch's limited register width.
Since multiple directories may have the same fingerprint,
{\sys} ensures that all directories with the same fingerprint (called a \emph{fingerprint group}) are partitioned to the same server, to simplify conflict handling.

\section{{\sys} Design}
In this section, we present how {\sys} works, including asynchronous metadata operations (\autoref{sec:design-state} and \autoref{sec:design-asynchronous-metadata-operations}), change-log compaction (\autoref{sec:design-aggregation}) and fault handling (\autoref{sec:design-fault-tolerance}).
We discuss the correctness of crash recovery (\autoref{sec:appendix-crash}) and the consistency properties (\autoref{sec:appendix-correctness}) in the appendix.
\label{sec:design}

\subsection{Directory State and Transition}
\label{sec:design-state}

\begin{figure}[t]
    \centering
    \includegraphics[width=0.75\linewidth]{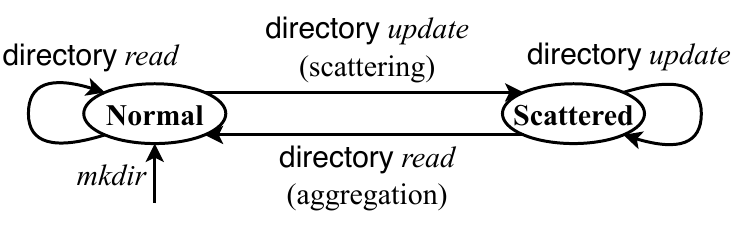}
    \vspace{-7pt}
    \caption{\textbf{Directory state transition.}}
    \label{fig:design-state-transition}
\end{figure}

\paragraph{State definition.} In {\sys}, %
a directory is in one of two states: \emph{normal state}, indicating that all returned updates to the directory have been applied to its inode, or \emph{scattered state}, indicating that there are not-yet-applied updates in one or more change-logs.
The on-switch dirty set tracks directory states, and metadata operations can query or update it.

\paragraph{State transition.} \autoref{fig:design-state-transition} shows the transitions between directory states.
A directory is created in normal state, where its inode is up-to-date.
When {\sys} makes an asynchronous update, the directory transitions to scattered state, indicating that there is one or more delayed updates not yet applied to the inode.
When {\sys} reads a directory in scattered state, it aggregates and applies the delayed updates from other servers, returning the directory to normal state.

\paragraph{Transition granularity.}
As mentioned in \autoref{sec:metadata schema}, the switch identifies a directory by its fingerprint.
Therefore, the granularity of state transition is fingerprint group, i.e., all directories with a specific fingerprint.
A \emph{metadata aggregation} aggregates all directories in the target fingerprint group.
Since {\sys} places all directories in the same fingerprint group to the same server, this does not complicate aggregation.

\subsection{Asynchronous Metadata Operations}
\label{sec:design-asynchronous-metadata-operations}

\paragraph{Overview.} Metadata operations can be classified into three categories according to the number of inodes they access~\cite{SingularFS}.
We summarize how {\sys} handles them below.

\begin{itemize}[itemsep=0pt,leftmargin=1em]
\item \emph{Double-inode operations}, including \emph{create}, \emph{delete}, \emph{mkdir}, \emph{rmdir}, access inodes of the target object and its parent directory.
These operations defer the update to the parent directory and finalize execution on a single metadata server; we elaborate upon them in \autoref{sec:directory-update} and \autoref{sec:rmdir}.

\item \emph{Single-inode operations} access the inode of a file (e.g., \emph{open}, \emph{close}, \emph{stat}) or a directory (e.g., \emph{statdir}, \emph{readdir}).
The former are performed synchronously as in traditional DFSs like InfiniFS~\cite{Lv2022InfiniFS} and CFS~\cite{CFS}. We omit their explanation for brevity.
The latter read directory attributes, and need to check inode staleness and possibly trigger an aggregation; we discuss them in \autoref{sec:directory-read}.

\item \emph{Rename} is the most complex metadata operation, as it updates up to four inodes.
{\sys} ensures \emph{rename} consistency through distributed transactions, and avoids \emph{orphaned loops}~\cite{Bjorner2007,Lv2022InfiniFS,CFS} (i.e., two directories become each other's ancestor) by using a centralized rename coordinator, as in previous DFSs~\cite{Lv2022InfiniFS,CFS}.
Notably, if the source is a directory, {\sys} initiates an aggregation at the beginning of \emph{rename} to apply all delayed updates.
\end{itemize}

\paragraph{Protocol intuitions.}
The intuition of {\sys}'s asynchronous metadata operation is to split a double-inode operation into two halves: the \emph{local half}, which updates the target object's inode and persists the update to the parent directory in a change-log, and the \emph{remote half}, which aggregates and applies the change-log entries to the parent directory's inode when it is read.
This effectively defers the directory update until read, allowing operations to return early and to hide cross-server overhead.

To ensure that asynchronous operations do not lose updates, {\sys} uses the on-switch dirty set to coordinate directory reads and writes.
We show that metadata operations are serializable and preserve real-time order in \autoref{sec:appendix-correctness}.

\paragraph{Data structures on metadata servers.}
A metadata server maintains the following data structures.

\begin{itemize}
    \item The \emph{change-log} is a per-server, per-directory FIFO queue that records the committed but not-yet-applied asynchronous updates to a directory.

    \item The \emph{invalidation list} is used for \emph{lazily} invalidating stale client metadata caches, as in InfiniFS~\cite{Lv2022InfiniFS}.
    When a directory is removed, renamed, or changes permission, the server that hosts the directory's inode broadcasts a \emph{message} to all servers, and the receivers append the directory's information to their invalidation lists.
    When contacting a server, a client consults the invalidation list and invalidates any stale cache entries accordingly.

    \item The \emph{key-value store} stores inodes and directory entries.
 
    \item The \emph{write-ahead-log} (WAL) is a persistent structure for crash recovery.
    It records the sequence of committed operations, and marks whether the corresponding asynchronous updates have been applied to remote servers.
\end{itemize}

The \emph{change-log}, \emph{invalidation list}, and \emph{key-value store} are kept in memory for performance.
\autoref{sec:design-crash-recovery} and \autoref{sec:appendix-crash} discuss how to recover them after crashes.

\begin{figure}[t]
    \centering
    \includegraphics[width=\linewidth]{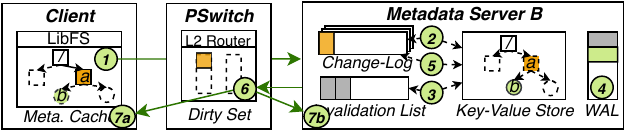}
    \vspace{-15pt}
    \caption{\textbf{Workflow of \emph{mkdir}, \emph{create} and \emph{delete}.}} %
    \label{fig:design-workflow-1}
\end{figure}

\subsubsection{
    Workflow of \emph{mkdir}, \emph{create} and \emph{delete}.}
\label{sec:directory-update}
These operations asynchronously update the parent directory's metadata.
\autoref{fig:design-workflow-1} illustrates an example of \emph{create(/a/b)}.
Other operations are handled similarly.

\paragraph{Path resolution.}
First, the client looks up each intermediate directory (i.e., \emph{/} and \emph{/a}) in its cache (\ding{192}).
On a cache miss, the client sends a \emph{lookup} request to the owner server of the directory's inode and fills the cache with inode information.
After the path resolution, the client sends the request to Server B, the owner server of \emph{/a/b}'s inode, which is determined by hashing \emph{/a/b}'s pid and name.

\paragraph{Locking and checking.}
Server B acquires a write lock on directory \emph{/a}'s change-log and a write lock on \emph{/a/b}'s inode in the key-value store (\ding{193}).
It then checks whether any path component of \emph{/a/b} appears in the invalidation list, and whether file \emph{/a/b} already exists (\ding{194}).
If invalid, Server B asks the client to invalidate stale cache entries and to retry the entire operation.
If the checks pass, the operation is persistently logged in the WAL and commits(\ding{195}).

\paragraph{Execution.}
Server B inserts the inode of file \emph{/a/b} into the key-value store and appends an entry to the change-log of directory \emph{/a} to record the \emph{create(/a/b)} operation (\ding{196}).

\paragraph{Dirty set update, reply, and unlocking.}
Server B sends to the switch a packet that contains the fingerprint of the parent directory \emph{/a}.
Upon receiving the packet, the switch inserts \emph{/a}'s fingerprint into the dirty set (\ding{197}).
If insertion succeeds, the switch multicasts the packet to both the client and Server B, notifying the client of the operation's completion (\Circled{7a}) and signaling Server B to release the locks (\Circled{7b}).
If the insertion fails (e.g., due to overflow), the operation falls back to synchronous metadata update: the switch forwards the packet to the server that owns the parent directory's inode, which then updates the parent directory synchronously.

\paragraph{Discussion.}
Notably, multiple servers may update the same directory and log their updates independently.
These updates do not conflict because updates to directory attributes are commutative, and the next \emph{statdir} or \emph{readdir} operation on the directory will aggregate all these updates (\autoref{sec:directory-read} and \autoref{sec:design-aggregation}).

\begin{figure}[t]
    \centering
    \includegraphics[width=\linewidth]{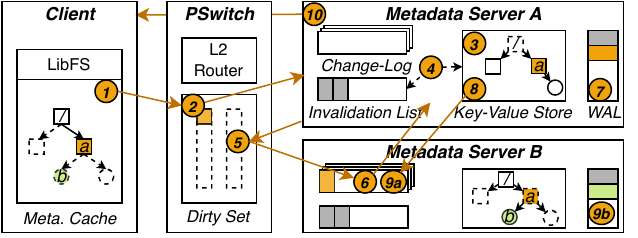}
    \vspace{-15pt}
    \caption{\textbf{Workflow of \emph{statdir}.}} %
    \label{fig:design-workflow-2}
\end{figure}

\begin{figure}[t]
    \centering
    \includegraphics[width=\linewidth]{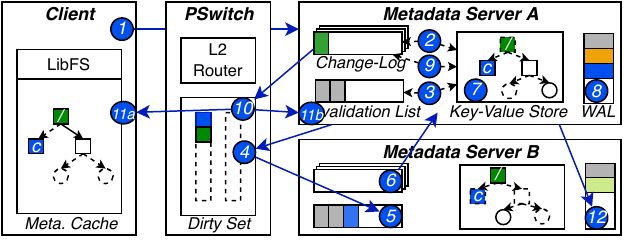}
    \vspace{-15pt}
    \caption{\textbf{Workflow of \emph{rmdir}.}} %
    \label{fig:design-workflow-3}
\end{figure}

\subsubsection{
    Workflow of \emph{statdir} and \emph{readdir}.}
\label{sec:directory-read}
These operations read directory attributes and entry lists, triggering aggregation if the directory is in \emph{scattered state}.
\autoref{fig:design-workflow-2} illustrates an example of \emph{statdir(/a)}.
\emph{readdir} is handled similarly.
\paragraph{Path resolution.}
The client performs path resolution as in \autoref{sec:directory-update} (\ding{192}), then sends the request packet to directory \emph{/a}'s owner server (i.e., Server A). This packet contains the fingerprint of \emph{/a}.

\paragraph{Dirty set query.}
When the switch forwards the packet, it queries \emph{/a}'s state and attaches the result to the packet (\ding{193}).

\paragraph{Locking and checking.}
Upon receiving the packet, Server A first acquires a read lock on \emph{/a}'s inode (\ding{194}) and checks cache validation and inode existence as in \autoref{sec:directory-update} (\ding{195}).
Then, the server examines the dirty set query result.
If \emph{/a} is in normal state, the server directly returns the directory's inode to the client.
Otherwise, Server A issues a metadata aggregation to renew the directory's inode before replying to the client.

\paragraph{Aggregation and reply.}
To issue an aggregation, Server A blocks \emph{statdir} and \emph{readdir} for all directories in \emph{/a}'s fingerprint group, and sends an aggregation request containing \emph{/a}'s fingerprint to the switch.
When the switch receives the request, it removes \emph{/a}'s \emph{fingerprint} from the dirty set and multicasts the request to all other servers (\ding{196}).
Upon receipt of the request, each other server acquires a read lock on \emph{/a}'s change-log, then sends all the change-log entries in \emph{/a}'s fingerprint group to Server A (\ding{197}).
When Server A receives these change-log entries, it logs each entry in the WAL (\ding{198}) and updates \emph{/a}'s metadata in the key-value store accordingly (\ding{199}).
After receiving entries from all servers and applying them, Server A multicasts an acknowledgment to the other servers, releases local locks,
and responds to the client (\ding{201}).
When other servers receive the acknowledgment, they unlock the change-logs (\Circled{9a}) and mark the change-log entries as ``applied'' in the WAL (\Circled{9b}).
The entries in the change-logs are reclaimed in the background.

\subsubsection{Workflow of \emph{rmdir}.}
\label{sec:rmdir}
\autoref{fig:design-workflow-3} illustrate an example of \emph{rmdir(/c)}. 
The first few steps, i.e., path resolution (\ding{192}), locking (\ding{193}), validation, and inode existence checks (\ding{194}) are the same as \emph{create}.
After these steps, Server A (i.e., the owner server of \emph{/c}) must collect the latest updates to \emph{/c} to determine whether the directory is empty, and prevent any stale client-side cache of \emph{/c} from being reused in future operations.
To do so, Server A sends an aggregation request to the switch.
The switch removes \emph{/a}'s fingerprint from the dirty set and multicasts the request to all other servers (\ding{195}).
Upon receiving the request, each other server (for example, Server B) 
inserts \emph{/c} into its invalidation list (\ding{196}), so that clients' stale cache of \emph{/c} will be invalidated during clients' next operations.
Then, Server B acquires a read lock on \emph{/c}'s change-log and sends the entries to Server A (\ding{197}).
Server A applies all received change-log entries and then checks the emptiness of \emph{/c} based on the aggregated metadata (\ding{198}).
If \emph{/c} is empty, Server A logs the \emph{rmdir(/c)} operation in the WAL (\ding{199}) and the operation commits.
Otherwise, the \emph{rmdir(/c)} fails with an \emph{ENOTEMPTY} error.
The remaining steps (\ding{200} \ding{201}, \Circled{11a}, \Circled{11b}) are similar to \emph{create}.
At the end, Server A sends a packet notifying other servers to release the locks and to mark the change-log entries they sent as ``applied'' in their WAL (\Circled{12}).

\paragraph{Discussion.}
{\sys} prevents clients from accessing stale directory metadata through lazy invalidation.
Specifically, \emph{rmdir(/c)} appends \emph{/c} to the invalidation lists on all servers (\ding{196}) before actually removing the directory (\ding{200}).
Any operation under \emph{/c} that checks path validity after the append fails the check and issues a \emph{lookup(/c)} request.
Since \emph{rmdir(/c)} holds a write lock on \emph{/c}'s inode (\ding{193}), the lookup waits until \emph{rmdir(/c)} completes, observing the latest metadata.

\subsection{Change-Log Compaction and Applying}
\label{sec:design-aggregation}

\begin{figure}[t]
    \centering
    \includegraphics[width=\linewidth]{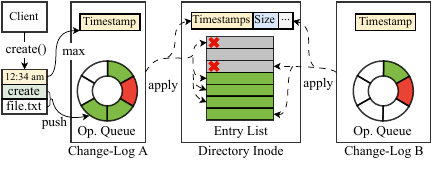}
    \vspace{-20pt}
    \caption{\textbf{Change-log compaction and application.}}
    \label{fig:design-changelog-apply}
\end{figure}

Asynchronous metadata updates enable parallel and local logging of directory modifications, hiding directory update overhead.
{\sys} further amortizes the update overhead by compacting change-logs before applying them to directory inodes, alleviating the metadata contention bottleneck.

\paragraph{Change-log structure.}
In {\sys}, a server maintains a local change-log for every \emph{scattered} directory.
A change-log entry represents a delayed directory update, containing the timestamp, operation type (\emph{create}, \emph{delete}, etc.), and filename, as depicted in \autoref{fig:design-changelog-apply}.

\paragraph{Opportunities for change-log compaction.}
Updates to the same directory are conditionally commutative, allowing merging multiple updates into a single one before applying them, mitigating update contention~\cite{180269,CFS,10.1145/3068914,10.1145/2757667.2757683,10.1145/3447865.3457970,10.1145/3465332.3470872,ahmednacer2012filecrdt}.
Specifically, a directory update involves three types of actions:
\begin{enumerate}[label=(\alph*), leftmargin=1.5em]
    \item apply a delta to attributes such as \emph{size},
    \item overwrite timestamps such as \emph{atime} and \emph{mtime}, and
    \item insert or remove an entry in the entry list.
\end{enumerate}

For type (a), applying the deltas in any order results in the same final state.
For type (b), only the largest timestamp remains.
For type (c), insertions and removals of different entries do not conflict and may be applied in any order, whereas repeated insertions and removals of the same name must be applied in their commit order (e.g., \emph{mkdir(/a), rmdir(/a)}).
In {\sys}, this order is preserved by the change-log's FIFO structure, given that insertions and removals of the same entry are always logged by the same server.

\paragraph{Change-log compaction workflow.}
Therefore, {\sys} proposes \emph{change-log compaction} to decrease the cost of directory update.
When directory updates are appended to a change-log, the change-log keeps the maximum timestamp of all entries, and stores the operation type and the filename.
Upon receiving an aggregation request, the server transmits the relevant change-log to the directory's owner server. The owner server then traverses the operation queue, updating the directory entry list, and updates the inode's attributes (e.g., timestamps and size) atomically.

Change-log compaction mitigates directory update contention in two ways.
First, asynchronous metadata updates allow all servers to consolidate updates to remote parent directories locally in their change-logs, enhancing inter-server parallelism.
Second, the change-log compaction consolidates timestamps and size deltas beforehand to minimize the number of key-value store's \emph{put()} operations.%

\paragraph{Proactive aggregation.}
To prevent a prolonged aggregation stall of the first directory read after a sequence of directory updates, a server proactively pushes its change-log entries to the directory's owner server (1) when the accumulated entries are sufficient to fill a maximum transmission unit (MTU), or (2) when no new entries have been appended to the change-log within a configured interval.
Upon receiving a change-log push, the directory's owner server starts a timer.
If no new entry pushes arrive within a configured period, the owner server proactively initiates an aggregation.
This proactive aggregation turns the directory back to normal state, allowing subsequent directory reads to complete without triggering an aggregation.

\subsection{Fault Tolerance}
\label{sec:design-fault-tolerance}

\subsubsection{Unreliable Network.}
\label{sec:design-fault-tolerance-network}
{\sys} protocol is based on UDP, which can experience packet loss and reordering.
{\sys} handles packet loss with timeout and resending.
The reordering of in-flight packets causes no issue, as packets belonging to the same operation are sent and received one by one, while packets belonging to different operations are independent, and {\sys} does not assume their order.%

Resending causes packet duplication.
Servers and the switch tolerate it with different mechanisms.
A server detects and drops duplicate packets by examining the (\emph{sender server, sequence number}) tuple attached to each packet by the sender server.
The switch handles three types of dirty set operations encapsulated in packets: \emph{fingerprint} \emph{query}, \emph{insert}, and \emph{remove}.
The switch executes duplicate queries and inserts without special treatment since they do not affect correctness.
Specifically, queries have no side effects, and redundant inserts may trigger unnecessary aggregations, but do not compromise correctness.
Only duplicate \emph{remove}s can cause inconsistency, in the case where a duplicate \emph{remove} request sent by an aggregation reaches the switch after the aggregation completes, removing fingerprints inserted by subsequent operations.
To avoid this, each \emph{remove} request has a sequence number (distinct from the packet's sequence number), which increments with every new request or resend.
The switch records the highest \emph{remove} sequence number received from each server and processes a \emph{remove} request only if its sequence number exceeds all previously received ones from the sending server.
After resending a \emph{remove} request, the server waits for responses corresponding to the last resend.
In this way, no duplicate \emph{remove} requests take effect after the aggregation completes, ensuring consistency.

\subsubsection{Crash Recovery}
\label{sec:design-crash-recovery}
\paragraph{Server failure.}
The servers maintain data structures in DRAM and use the write-ahead log (WAL) to recover lost progress after server failures, as production deployments (e.g., PanguFS and HDFS) typically do~\cite{Lv2022InfiniFS}.
To recover from failures, the server first redoes operations in the WAL to rebuild its local key-value store and change-log entries that haven't been marked as ``applied''.
Note that we don't need to rebuild change-log entries marked as ``applied'', as they have been persisted in the WAL of the directory inode's server.
Then, the server proactively aggregates all directories it owns to ensure any interrupted aggregations it issued before the crash are executed to completion.
Finally, the server recovers its invalidation list by cloning it from other servers.
During recovery, the server does not serve normal requests.

\paragraph{Switch failure.}
After a switch failure, all state in the switch are lost.
{\sys} initializes an empty dirty set in the switch and notifies all servers to aggregate all directories.
Once all aggregations are complete, all change-log entries have been sent to and applied by their directory inode's server.
Thus, all directories in {\sys} return to normal state, consistent with the empty dirty set.
During recovery, all servers stop serving normal requests.

\subsection{Discussion}
\label{sec:discussion}

\paragraph{Cluster reconfiguration.}
{\sys} uses consistent hashing~\cite{consistenthashing} to map inodes to servers.
When servers are added or removed, a small portion of metadata needs to be migrated.
{\sys} performs migration in a stop-the-world manner:
all metadata servers stop serving requests, aggregate all directories, and then migrate the metadata to the new server via two-phase commit.
Since the hash function resides on the clients and servers, no changes are needed on the switch.
We discuss the fault tolerance of cluster reconfiguration in \autoref{sec:appendix-reconfiguration}.

\paragraph{Support of hard links.}
To support hard links, the file metadata in \autoref{tab:metadata schema} needs to be split into two KV objects:
\begin{itemize}
    \item the \emph{reference}: (pid, name) $\rightarrow$ (server\_id, file\_id), and
    \item the \emph{attributes}: file\_id $\rightarrow$ (ref\_count, size, timestamps, \dots).
\end{itemize}
On file creation, both objects are created on the server determined by hashing \emph{(pid, name)}.
When a hard link is created, only a new reference is created at the hashing location, pointing to the original file's attributes, which may reside on a different server.
The attributes' \emph{ref\_count} tracks the number of references, and the attributes are deleted when it drops to zero.
For consistency and atomicity, both objects are protected by two-phase locking during metadata access, and two-phase commit coordinates updates across servers.

\section{Data Plane Design}
\label{sec:design_of_switch_data_plane}

In this section, we introduce the format of {\sys} packets (\autoref{sec:dataplane-packetformat}), the switch data plane layout (\autoref{sec:dataplane-layout}), and the dirty set design (\autoref{sec:dataplane-staleset}). We also discuss scaling to multiple racks (\autoref{sec:dataplane-deployment}) and the adaptability issues (\autoref{sec:data-plane-discussion}).

    \subsection{{\sys} Packet Format}
    \label{sec:dataplane-packetformat}

    {\sys} employs UDP as the transport layer protocol for lightweight networking.
    As illustrated in \autoref{fig:dataplane-filter}, the UDP payload begins with an optional header that encapsulates a dirty-set operation, which can be parsed by the switch.
    Following this header, the payload contains a DFS request or response for servers to process.
    {\sys} reserves two UDP ports for {\sys} packets with and without the dirty-set operation header so that the switch can distinguish them.

    \subsection{Data Plane Layout}
    \label{sec:dataplane-layout}

    \begin{figure}[t]
        \centering
        \includegraphics[width=\linewidth]{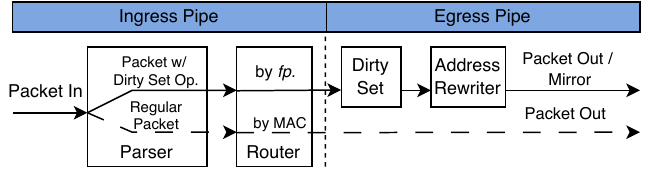}
        \vspace{-15pt}
        \caption{\textbf{Logical view of {\sys} switch data plane.}}
        \label{fig:data_plane}
    \end{figure}

    Figure~\ref{fig:data_plane} presents the layout of the switch data plane, which consists of the following components.

    \paragraph{Parser.}
    The \emph{parser} parses the packet headers to extract the fields for subsequent processing.

    \paragraph{Router.} 
    The \emph{Router} sets the egress port field in the switch's metadata for (a) regular packets, by destination MAC address, and (b) packets with dirty-set operations, by \emph{fingerprint} prefix. The switch forwards the packets accordingly.

    \paragraph{Dirty set.}
    The \emph{dirty set} is implemented as a multi-slot hash table that supports three kinds of operations:
    (1) \emph{insert} a fingerprint into the set;
    (2) \emph{query} whether a fingerprint exists in the set;
    (3) \emph{remove} a fingerprint from the set.
    When a packet arrives, the dirty set executes the operation specified in the \emph{OP} field of the header and writes the result to the \emph{RET} field.
    \emph{SEQ} is a sequence number for recognizing duplicate \emph{remove} requests, as described in \autoref{sec:design-fault-tolerance-network}.

    Since a switch may have multiple pipes and pipes do not share state, we place the dirty set in the egress pipes in a shared-nothing manner.
    Each pipe is responsible for fingerprints with different prefixes, and packets are routed accordingly.
    If a packet accesses a fingerprint managed by a pipe different from its destination port, the switch mirrors it to the destination pipe, as in prior studies~\cite{Jin2017NetCache,Yu2020NetLock}.

    \paragraph{Address rewriter.}
    If a dirty-set insertion fails due to overflow, the switch redirects the packet using the \emph{alternative MAC address} field in the packet header for fallback handling.
    The \emph{address rewriter} overwrites the L2 destination field with this alternative address, and the switch forwards accordingly.
    
    \subsection{In-Network Dirty Set}
    \label{sec:dataplane-staleset}

    The on-switch dirty set maintains the set of directories in scattered states.
    Since its performance and capacity are critical to {\sys}, 
    we have the following design goals:
    (1) support line-rate \emph{query}, \emph{insert} and \emph{remove} of fingerprint, and
    (2) maximize the utilization of scarce on-chip memory.

        \begin{figure}[t]
            \centering
            \includegraphics[width=\linewidth]{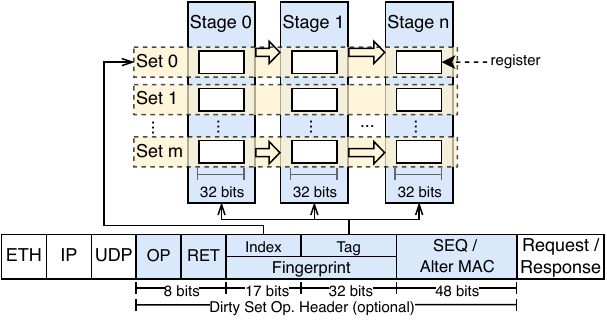}
            \vspace{-15pt}
            \caption{\textbf{Data structure of the dirty set.}}
            \label{fig:dataplane-filter}
        \end{figure}

        \begin{figure}[tb]
            \centering
            \includegraphics[width=\linewidth]{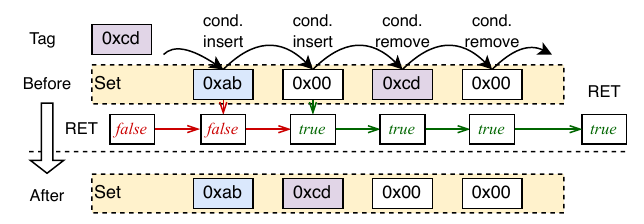}
            \vspace{-15pt}
            \caption{\textbf{An example of the dirty-set \emph{insert} operation.}}
            \label{fig:dataplane-operation}
        \end{figure}

        \paragraph{Structure.}

        Due to programming constraints that prevent the implementation of complex indexing structures,
{\sys} organizes 32-bit registers in a manner similar to a set-associative cache to store fingerprints.
        As \autoref{fig:dataplane-filter} shows, the switch data plane comprises multiple stages, each containing an array of registers.
        Registers at the same position across stages are horizontally grouped into a set.
        In our switch configuration, there are ten stages, and each stage allocates 131,072 ($2^{17}$) registers.
        This setup allows the switch to store up to 1,310,720 fingerprints. We use the upper 17 bits of the fingerprint as the set \emph{index}, while the remaining 32 bits serve as the \emph{tag} to uniquely identify each fingerprint.

        \paragraph{Dirty-set operations.}

        The dirty set supports three operations: \emph{insert}, \emph{query}, and \emph{remove}.
        Each operation performs a sequence of \emph{register actions} on the registers indexed by the header's index field.
        We define three register actions:
        (a) \emph{register query} compares the register's value with \emph{tag} and returns the result;
        (b) \emph{conditional insert} returns whether the the register's 
        value equals zero or \emph{tag} and writes \emph{tag} into the register if the old value is zero;
        (c) \emph{conditional remove} writes zero into the register if the old value matches \emph{tag}.

For dirty-set \emph{query}, all stages perform register query sequentially, and it returns \emph{true} if any of the stages return \emph{true}. 
For dirty-set \emph{remove}, all stages perform conditional remove sequentially.
For dirty-set \emph{insert}, stages perform conditional insert one by one until one of them returns \emph{true}, and the following stages perform \emph{conditional remove} to ensure no duplicate \emph{tags} remain in the set.
\autoref{fig:dataplane-operation} illustrates an example of dirty set \emph{insert}.
    
        \paragraph{Properties.}
        The switch architecture provides two properties~\cite{264822}.
        (a) \emph{Atomicity.} Operations in the same stage are atomic.
        (b) \emph{Ordered execution.} For any two packets $P_A, P_B$ and two ordered stages $S_A, S_B$, if $P_A$ reaches $S_A$ before $P_B$, then $P_A$ reaches $S_B$ before $P_B$.

        Based on these properties, our design ensures that the dirty-set operations are
        (a) \emph{idempotent}, since successive repeated operations have the same effects as a single one, and
        (b) \emph{linearizable}, since operations on the same \emph{fingerprint} are serialized by the pipeline.

    \begin{figure}[tb]
        \centering
        \includegraphics[width=\linewidth]{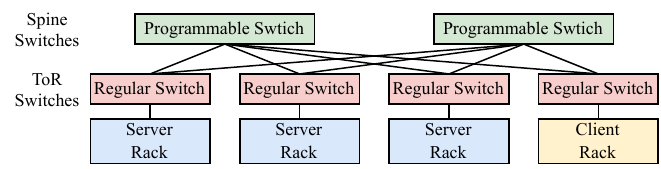}
        \vspace{-15pt}
        \caption{\textbf{Topology of the multi-rack deployment.}}
        \label{fig:dataplane-topology}
    \end{figure}

\subsection{Scale to Multiple Racks}
\label{sec:dataplane-deployment}
For single-rack deployment, where all metadata servers reside in the same rack, the dirty set is placed in the top-of-rack (ToR) programmable switch, enabling it to monitor all rack traffic.
For multi-rack deployments, {\sys} uses a leaf-spine topology as shown in \autoref{fig:dataplane-topology}.
Since ToR switches no longer have a global view, programmable switches are deployed at the spine layer.
To scale further, {\sys} range-partitions the state management of directories across switches by fingerprint, and directs packets involving each subset of directories to the corresponding switch.
A previous study~\cite{Harmonia} has shown that the overhead of the extra network hops between switches is negligible, since it is orders of magnitude smaller than that of operation processing.
This approach adapts to skewed workloads since (a) the distribution of directories over switches is uniform due to random fingerprints, and (b) each switch can process all the traffic in the metadata cluster without becoming a bottleneck (\autoref{eval:Contribution pswitch}).

\subsection{Discussion}
\label{sec:data-plane-discussion}
\paragraph{Adaptability of the programmable switch.}
{\sys}'s on-switch program comprises 847 lines of P4 code and 303 lines of C code.
It consumes 1,310,720 32-bit on-chip registers (5\;MiB in total).
Initializing it on a Tofino switch~\cite{Tofino} takes less than ten seconds.

\paragraph{Limitations.}
Given the limited on-chip memory of the Tofino switch~\cite{Tofino}, colocating {\sys} with state-heavy network functions (e.g., in-network caching~\cite{Jin2017NetCache,Liu2019DistCache}) is challenging.
However, it can coexist with lightweight functions such as flow control and packet filtering.
The programmable switch also constitutes a potential single point of failure, which we leave for future work.

\section{Evaluation}

\subsection{Experimental Setup}

    \paragraph{Hardware configuration.}
We conduct experiments with eight server machines and three client machines.
\autoref{tab:hardware configurations} shows their specifications.
    All machines are connected via a Wedge100BF-32X programmable switch equipped with a Tofino ASIC~\cite{Tofino}.
    To scale up the evaluation to 16 servers, we deploy a metadata server on each socket of the server nodes, utilizing the NUMA-local cores, DRAM, and NIC.
    We do not evaluate with multiple programmable switches due to hardware limitations.

\begin{table}[t]
\centering
\footnotesize
\begin{tabular}{@{}c|lllll@{}}
\toprule
\multirow{4}{*}{\begin{tabular}[c]{@{}c@{}}Server\\ Cluster\end{tabular}} & CPU     & \multicolumn{4}{l}{2 $\times$ Intel Xeon Gold 5317 3.00GHz, 12 cores}  \\
                                                                        & Memory  & \multicolumn{4}{l}{16 $\times$ DDR4 2933MHz 16GB}                      \\
                                                                        & Storage & \multicolumn{4}{l}{Intel Optane Persistent Memory}                     \\
                                                                        & Network & \multicolumn{4}{l}{2 $\times$ ConnectX-5 Single-Port 100GbE}           \\ \midrule
\multirow{3}{*}{\begin{tabular}[c]{@{}c@{}}Client\\ Cluster\end{tabular}} & CPU     & \multicolumn{4}{l}{2 $\times$ Intel Xeon E5-2650 v4 2.20GHz, 12 cores} \\
                                                                        & Memory  & \multicolumn{4}{l}{16 $\times$ DDR4 2933MHz 16GB}                      \\
                                                                        & Network & \multicolumn{4}{l}{2 $\times$ ConnectX-4 Single-Port 100GbE}           \\ \bottomrule
\end{tabular}
\caption{\textbf{Hardware specifications of clusters.}}
\label{tab:hardware configurations}
\end{table}

\paragraph{Baseline systems.}
We compare {\sys} with CephFS v12.2.13~\cite{Weil2006Ceph}, IndexFS~\cite{Ren2014IndexFS}, InfiniFS~\cite{Lv2022InfiniFS}, and CFS~\cite{CFS}.
CephFS is a widely used commercial DFS, while the others are state-of-the-art DFSs optimized for metadata performance.
Since InfiniFS and CFS are not publicly available, we implemented them from scratch.
We emulate InfiniFS by carefully implementing it according to its paper~\cite{Lv2022InfiniFS} and emulate CFS by replacing InfiniFS's P/C grouping approach (i.e., per-directory hashing) with CFS's P/C separation approach (i.e., per-file hashing).
Notably, Emulated-InfiniFS (E-InfiniFS), Emulated-CFS (E-CFS), and {\sys} share the same storage and networking framework, ensuring a fair comparison.
We do not compare with SingularFS~\cite{SingularFS} and DeltaFS~\cite{DeltaFS}, as SingularFS focuses on enhancing per-server performance with novel hardware (i.e., persistent memory and RDMA), which is orthogonal to our work, and DeltaFS is designed for HPC jobs, whose semantics differ from POSIX DFSs.

\paragraph{Configuration details.}
E-InfiniFS, E-CFS, and {\sys} use RocksDB in asynchronous write mode for metadata storage, and employ a coroutine-based, non-blocking RPC engine on DPDK 21.11.2~\cite{DPDK} for networking.
Unless specified otherwise, each metadata server uses four cores, and clients use LibFS to communicate with servers.
For CephFS, we deploy multiple active metadata server daemons (MDSs) colocated with object storage daemons (OSDs).
{\sys} enables proactive change-log aggregation in all experiments, so aggregation overhead is included in the results.
No dirty-set overflow occurs in the evaluation.

\subsection{Overall Performance}
\label{eval:Overall Performance}

\begin{figure*}[t]
\centering
\begin{subfigure}[b]{0.65\linewidth}
\centering
\includegraphics[width=\linewidth]{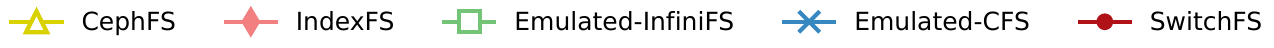}
\vspace{-15pt}
\end{subfigure}
\begin{subfigure}[b]{0.49\linewidth}
\centering
\includegraphics[width=\linewidth]{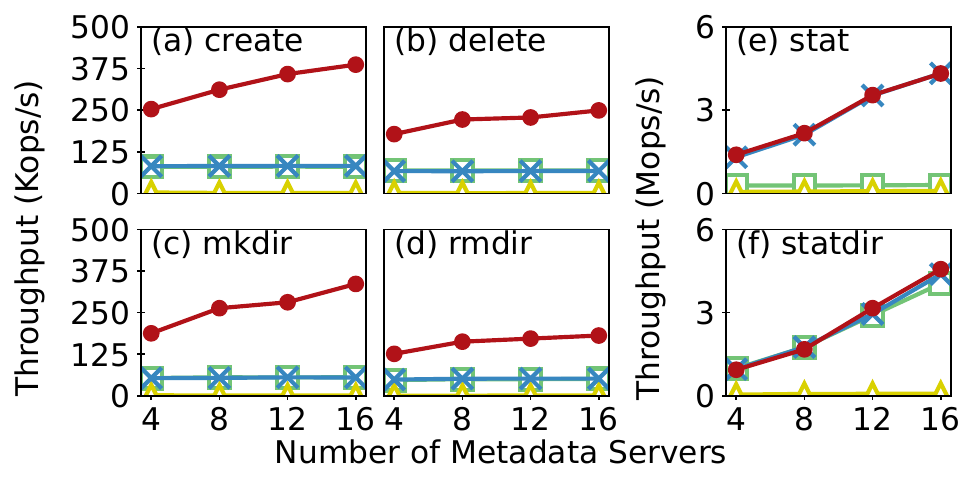}
\vspace{-15pt}
\caption{Single large directory.}
\label{fig:single throughput}
\end{subfigure}
\begin{subfigure}[b]{0.49\linewidth}
\centering
\includegraphics[width=\linewidth]{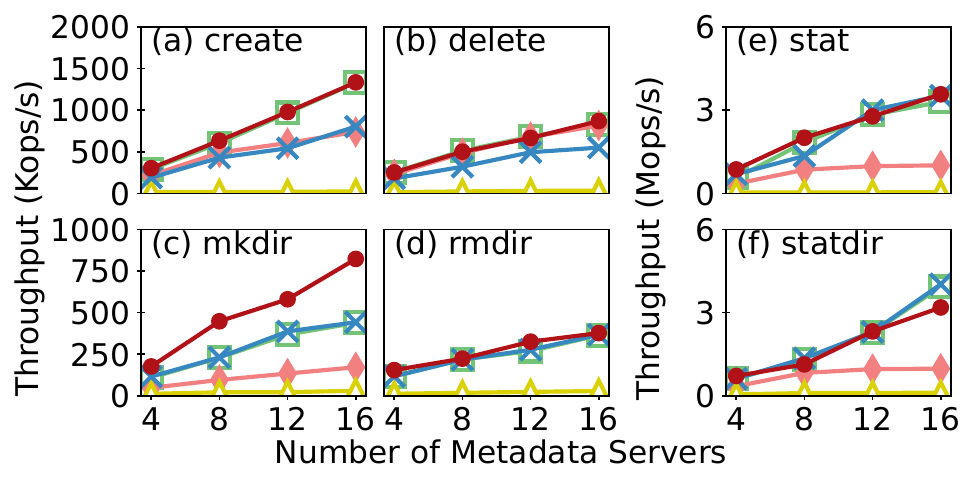}
\vspace{-15pt}
\caption{Multiple directories.}
\label{fig:rand throughput}
\end{subfigure}
\caption{\textbf{Throughput of metadata operations.}}
\label{fig:throughput}
\end{figure*}

\begin{figure}[t]
\centering
\includegraphics[width=\linewidth]{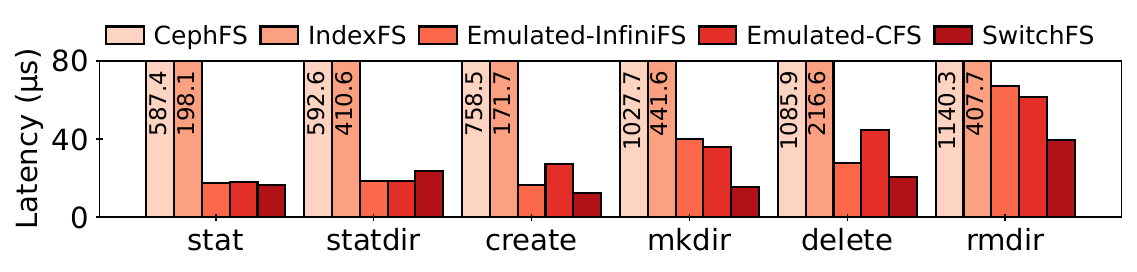}
\vspace{-20pt}
\caption{\textbf{Metadata operation latency.}}
\label{fig:latency}
\end{figure}

    \subsubsection{Throughput}
    We first evaluate how the peak throughput of individual operations scales with the number of servers.
    We gradually increase the number of concurrent requests issued by clients (up to 512) until the throughput no longer increases.
    We conduct experiments on two different access patterns, \emph{a single large directory} and \emph{multiple directories}, which reflect DFSs' load balance and operation overhead, respectively.
    IndexFS's throughput on a single large directory is not included since it consistently crashes with errors.

\paragraph{A single very large directory.}

In this evaluation, clients perform operations on 10 million files in a single large directory and randomly select files to access.
\autoref{fig:single throughput} shows the operation throughput as the metadata server increases. We make the following observations.
   
1) {\sys}'s throughput scales well for double-inode operations including \emph{create}, \emph{delete}, \emph{mkdir} and \emph{rmdir}, as server number increases.
The reason is that,
(a) {\sys}'s fine-grained partitioning approach distributes loads across servers evenly;
(b) the asynchronous metadata update protocol allows double-inode operations to access only one server, avoiding cross-server coordination and hiding directory contention;
(c) change-log compaction merges conflicting directory updates, reconciling the directory contention.
Note that {\sys}'s \emph{rmdir} throughput is lower than that of \emph{mkdir} since {\sys} uses multicast to notify the invalidation of the directory during \emph{rmdir}.

2) Despite employing fine-grained partitioning, E-CFS exhibits minimal scalability for \emph{create}, \emph{delete}, \emph{mkdir}, and \emph{rmdir}.
This arises because these operations are serialized per directory, leading to severe contention.

3) Both {\sys} and E-CFS scale linearly for \emph{stat} operations, benefiting from effective load balancing via fine-grained partitioning. 
In contrast, E-InfiniFS suffers from significant load imbalance due to assigning all files within a large directory to a single server.
For \emph{statdir}, all filesystems except CephFS scale linearly, as they partition directories in a balanced manner.
{\sys}'s dirty set checking introduces negligible additional overhead to \emph{statdir}.
   
4) %
CephFS's throughput is low (below 100 Kops/s) for its heavy software stack and inflexible subtree partitioning.

    \paragraph{Multiple directories.}

In this evaluation, clients randomly access 10 million files uniformly distributed across 1024 directories.
\autoref{fig:single throughput} shows the operation throughput as the number of metadata servers increases.
We make the following observations.

1) In this setting, operations have little contention and DFSs achieve optimal performance.
{\sys} performs the best for all operations, indicating that asynchronous metadata updates have negligible overhead on the common path. %

2) For \emph{create} and \emph{delete}, {\sys} and E-InfiniFS perform similarly and outperform E-CFS.
E-InfiniFS performs well as its parent-children grouping enables updating the file and parent directory metadata on the same server.
{\sys} achieves comparable performance since asynchronous metadata updating hides the overhead for updating directory metadata.
In contrast, E-CFS performs worse since it separates file and parent directory metadata, requiring cross-server transactions for updates.

3) For \emph{mkdir}, {\sys} outperforms IndexFS, E-InfiniFS, and E-CFS since asynchronous updates hide cross-server coordination overhead for updating two directories' metadata on different servers, while the other DFSs need to use distributed transactions.
For \emph{rmdir}, {\sys} performs similarly to E-InfiniFS and E-CFS, as the multicast overhead offsets the gain of asynchronous updates.
IndexFS's implementation of \emph{rmdir} is incomplete, so its results are omitted.

\begin{figure}[t]
\centering
\begin{subfigure}[b]{0.495\linewidth}
\centering
\includegraphics[width=\linewidth]{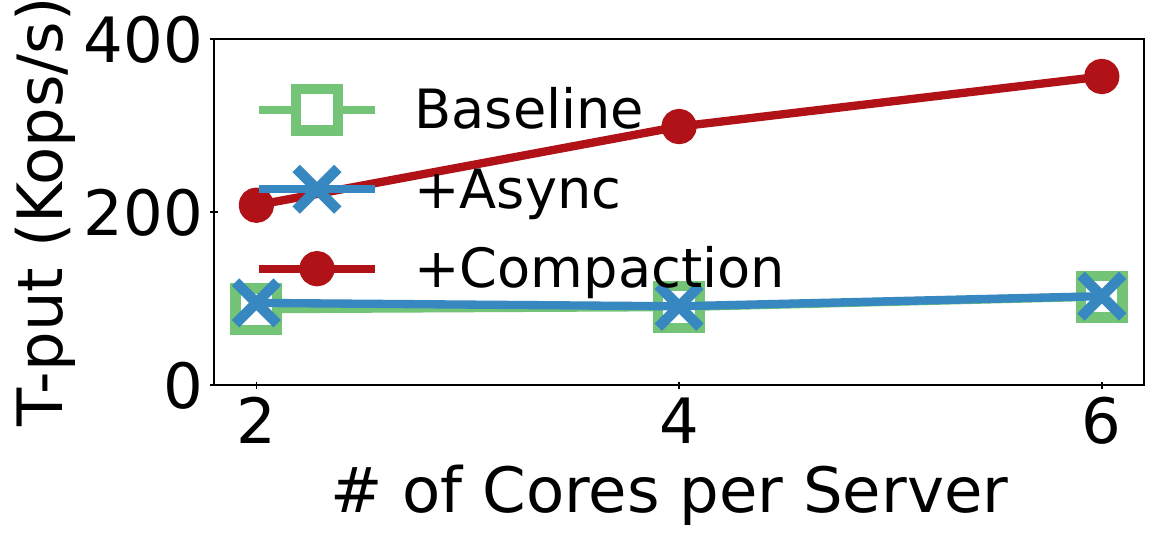}
\end{subfigure}
\begin{subfigure}[b]{0.495\linewidth}
\centering
\includegraphics[width=\linewidth]{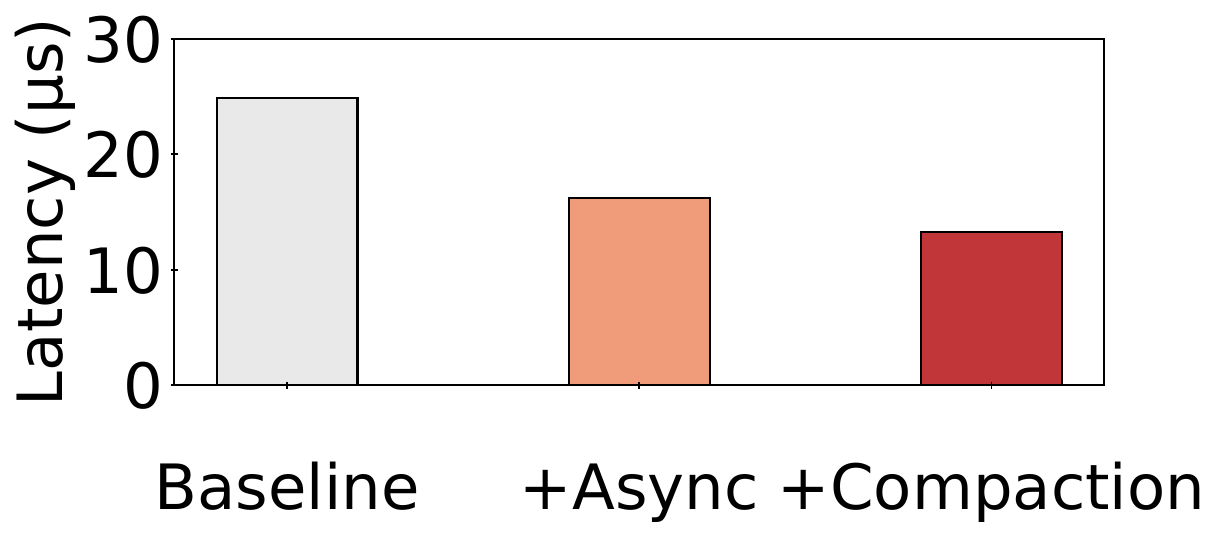}
\end{subfigure}
\vspace{-20pt}
\caption{\textbf{Contribution breakdown.} Performance of file {create} in a single directory with different techniques enabled.}
\label{fig:breakdown}
\end{figure}

\subsubsection{Latency}
We evaluate the latency of operations by using a single client to issue requests one by one. DFSs have eight servers.
\autoref{fig:latency} shows the average latency.

1) {\sys} significantly reduces the latency of \emph{create}, \emph{delete}, \emph{mkdir}, and \emph{rmdir} compared to other DFSs, because its asynchronous metadata updates hide the overhead for updating the parent directory.

2) {\sys}'s \emph{statdir} latency is 28.6\% higher than E-InfiniFS and E-CFS due to additional checks introduced by asynchronous metadata update for correctness.

\subsection{Contribution Breakdown}
\label{eval:Contribution Breakdown}

\subsubsection{Contribution of Techniques.}
\label{sec:contribution of techniques}
In this evaluation, we analyze how each design feature contributes to {\sys}'s throughput and latency.
Clients create 10 million files in a single directory, and the DFS has eight servers.
When evaluating throughput, we further vary the number of cores per server to demonstrate {\sys}'s intra-server parallelism.
\autoref{fig:breakdown} shows the result.

\textbf{Baseline} is the framework of {\sys}, employing fine-grained partitioning and synchronous operations.
It suffers from high latency due to cross-server coordination and low throughput caused by parent-directory contention.

\textbf{+Async} adopts asynchronous metadata updates without change-log compaction.
Compared to \emph{Baseline}, the average latency is reduced by 34.7\%, while the throughput remains unchanged.
This is because asynchronous metadata updates hide the latency of updating the parent directory, but do not reduce the updating overhead.
During aggregation, updates to the parent directory's metadata are serialized by the key-value store due to contention, resulting in poor parallelism.

\textbf{+Compaction} is the complete {\sys} design with both asynchronous metadata updates and change-log compaction optimization.
Compared to previous settings, the throughput is improved by up to 2.48$\times$ and scales as the number of cores per server increases.
This is because the change-log compaction merges multiple updates to the directories' attributes (e.g., timestamps) into a single one, and the directory entries are inserted or deleted in parallel.
Compared to \emph{+Async}, the average latency is reduced by 18.2\%, and the p99 latency drops from 173\,$\mu$s to 22\,$\mu$s.

\subsubsection{Impact of Dirty-Set Overflow.}
If insertion into the programmable switch's dirty set fails due to overflow, the operation falls back to server-side synchronous updates (\autoref{sec:directory-update}).
We force dirty set insertion to always fail and evaluate the performance of \emph{create} using the setup described in \autoref{sec:contribution of techniques}.
Compared to scenarios where insertions succeed, throughput drops by 69.7\% and average latency rises by 0.85$\times$, closely matching the performance of \emph{Baseline} in \autoref{sec:contribution of techniques}.

\begin{figure}[t]
    \centering
    \begin{subfigure}[b]{0.495\linewidth}
        \centering
        \includegraphics[width=\linewidth]{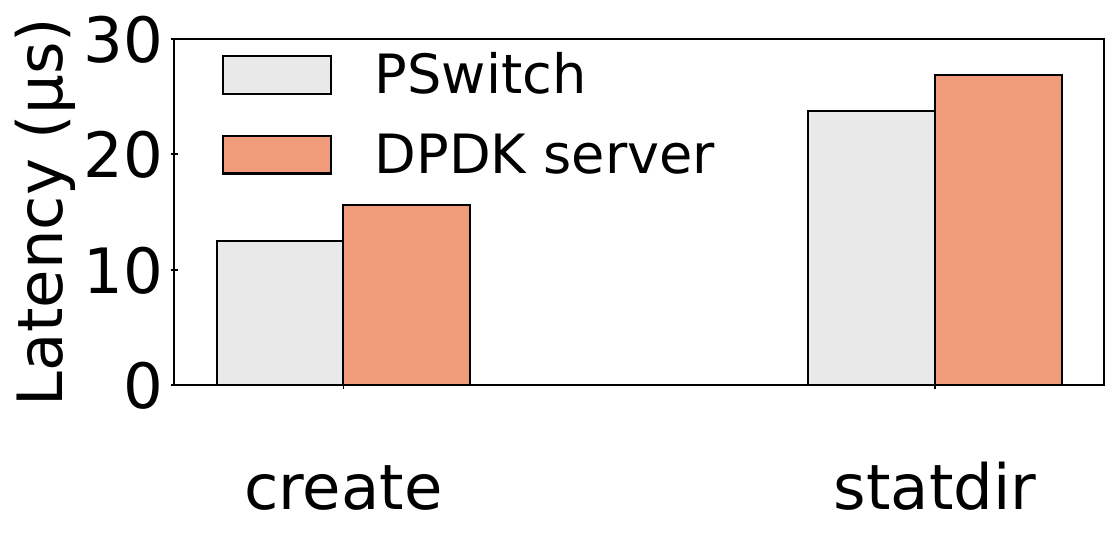}
        \vspace{-15pt}
        \caption{Latency of \emph{create}.}
        \label{fig:breakdown-switch latency}
    \end{subfigure}
    \begin{subfigure}[b]{0.495\linewidth}
        \centering
        \includegraphics[width=\linewidth]{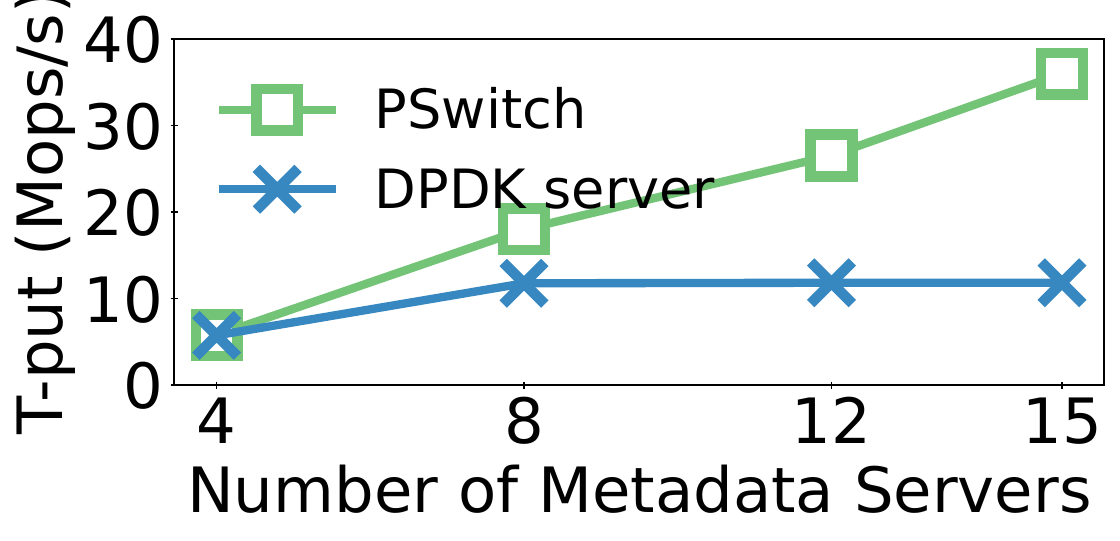}
        \vspace{-15pt}
        \caption{Throughput of \emph{statdir}.}
        \label{fig:breakdown-switch throughput}
    \end{subfigure}
    \caption{\textbf{Tracking directory states in a dedicated server.}}
    \label{fig:breakdown-switch-1}
\end{figure}

\begin{figure}[t]
    \centering
    \begin{subfigure}[b]{0.495\linewidth}
        \centering
        \includegraphics[width=\linewidth]{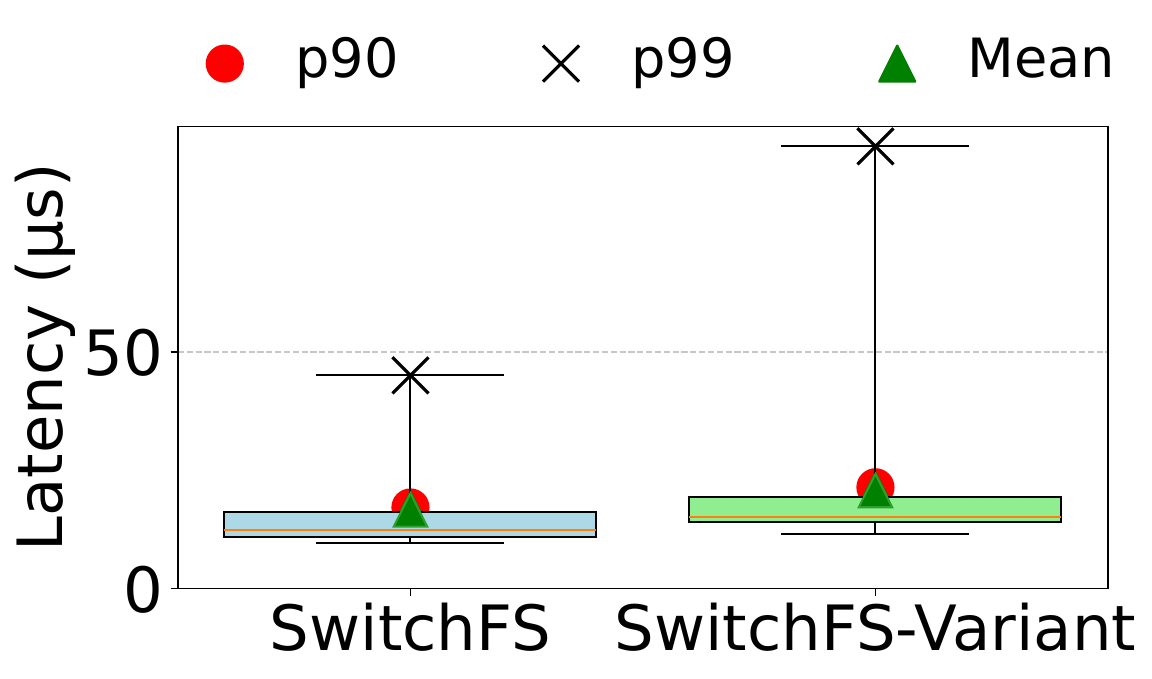}
        \vspace{-15pt}
        \caption{Under medium 50 Kops/s load.}
        \label{fig:latency-cdf-low-presure}
    \end{subfigure}
    \begin{subfigure}[b]{0.495\linewidth}
        \centering
        \includegraphics[width=\linewidth]{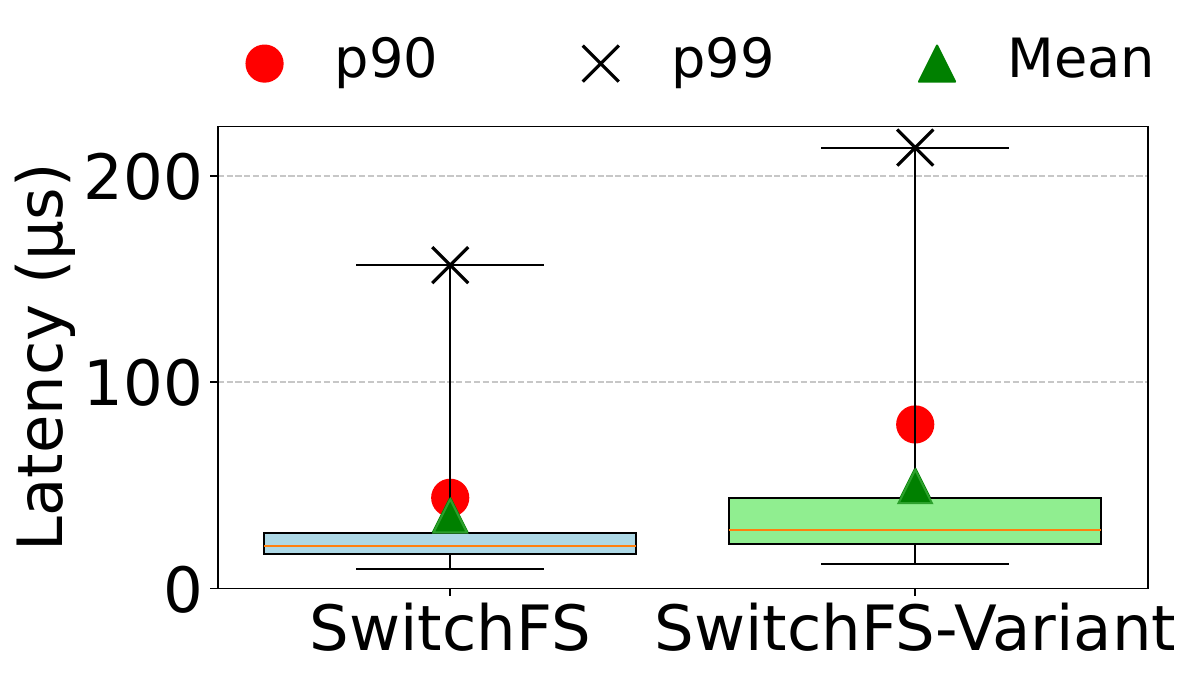}
        \vspace{-15pt}
        \caption{Under heavy 120 Kops/s load.}
        \label{fig:latency-cdf-high-presure}
    \end{subfigure}
    \caption{\textbf{Latency of \emph{create} operations when directory states are tracked on owner servers.}
    The top and bottom edges of each box represent p75 and p25, respectively, and the line in the box represents the median.
    }
    \label{fig:breakdown-switch-2}
\end{figure}

\subsubsection{Contribution of the Programmable Switch}
\label{eval:Contribution pswitch}
The asynchronous update protocol in {\sys} is not tightly coupled to the programmable switch.
This evaluation examines the trade-off of using a programmable switch versus two alternatives: 
(a) using a dedicated server to maintain the dirty set, and
(b) assigning each directory's owner server to track its own dirty state.

\paragraph{Tracking directory state with a dedicated server.}
Tracking directory state with a dedicated server, rather than a programmable switch, has two disadvantages.
First, as shown in \autoref{fig:breakdown-switch latency}, operations involving the dirty set incur an additional RTT ($\sim$3$\mu$s), raising the average latency of \emph{create} and \emph{statdir} by 24.1\% and 13.1\%, respectively.
Second, the server cannot match the throughput of a programmable switch.
In \autoref{fig:breakdown-switch throughput}, we measure throughput by performing \emph{statdir} on 1 million directories, with each server using 12 cores and the dedicated server leveraging DPDK~\cite{DPDK} for packet processing.
The dedicated server hits a throughput ceiling of 11 Mops/s, whereas the programmable switch scales linearly with the number of metadata servers.
Theoretically, a programmable switch can process up to 4.8 Bops/s~\cite{TofinoSpecifications}, far exceeding the demand of a metadata cluster.
Consequently, a single programmable switch suffices to serve the cluster, while dedicated servers would need to partition fingerprints across multiple machines to meet throughput demands.

\paragraph{Tracking directory state with the owner server.}
Maintaining directory state on the owner server instead of a programmable switch reduces the peak throughput of double-inode operations such as \emph{mkdir} and \emph{create} by $\sim 10\%$, as it doubles the number of packets processed per operation, consuming valuable CPU resources.
Since metadata performance is CPU-bound, this lowers overall throughput.

Moreover, tracking on owner servers substantially increases operation latency.
\autoref{fig:breakdown-switch-2} shows the latency distribution of \emph{create} under medium and heavy load.
Compared with the original {\sys}, this variant increases median, p90, and p99 latency by 23.2\%, 24.9\%, and 107.4\% under medium load, and by 36.9\%, 80.8\%, and 36.4\% under heavy load.
The extra server on the critical path of directory updates introduces additional queuing and head-of-line blocking, amplifying latency --- especially under high load.

\subsection{Operation Burst Performance}
\label{eval:Burst Performance}

In this evaluation, we evaluate how {\sys} handles temporal load imbalance; that is, a group of spatially related operations performed by applications within a short time period.
We model temporal load imbalance as operation bursts, defined as a group of successive file \emph{create}s in the same directory.
Successive operation bursts create files in different directories.
{\sys} has eight servers, and clients issue 32 or 256 in-flight requests to stress the servers.
We observe the throughput changes while varying the burst size.

\begin{figure}[t]
    \centering
    \begin{subfigure}[b]{\linewidth}
        \centering
        \includegraphics[width=0.8\linewidth]{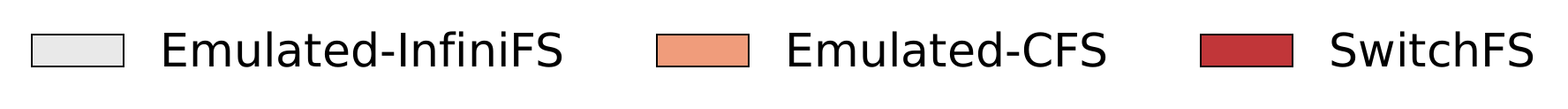}
        \vspace{-5pt}
    \end{subfigure}
    \begin{subfigure}[b]{0.495\linewidth}
        \centering
        \includegraphics[width=\linewidth]{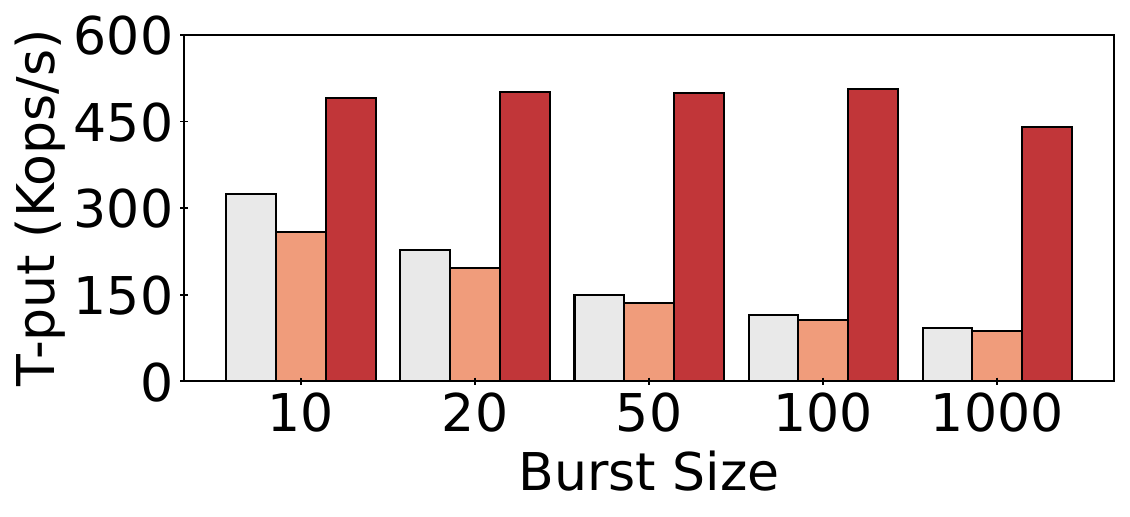}
        \vspace{-15pt}
        \caption{32 in-flight requests.}
        \label{fig:burst-32}
    \end{subfigure}
    \begin{subfigure}[b]{0.495\linewidth}
        \centering
        \includegraphics[width=\linewidth]{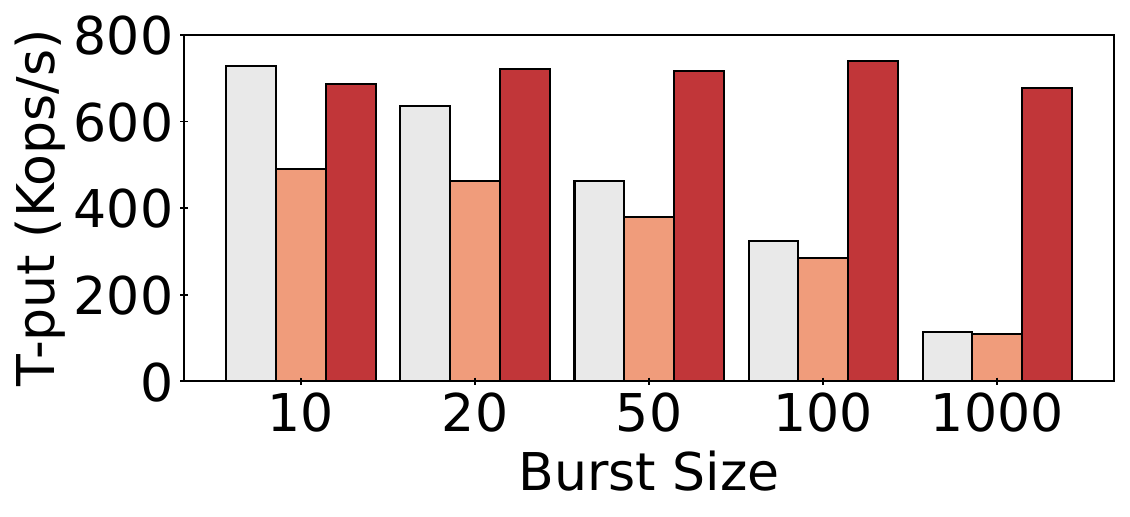}
        \vspace{-15pt}
        \caption{256 in-flight requests.}
        \label{fig:burst-256}
    \end{subfigure}
    \caption{\textbf{Throughput of \emph{create} in bursts.}}
    \label{fig:burst}
\end{figure}

As shown in \autoref{fig:burst}, both E-InfiniFS and E-CFS are sensitive to operation bursts.
With 32 in-flight requests (\autoref{fig:burst-32}), compared to burst size 10, the throughput of E-InfiniFS and E-CFS drops by 53.7\% and 47.1\% at burst size 50, and drops by 71.9\% and 66.1\% at burst size 1000.
In contrast, {\sys} exhibits stable performance as the burst size increases, because it buffers bursts in the change-log and applies them later.
\autoref{fig:burst-256} shows results with 256 in-flight requests, which imposes greater pressure on the servers.
Even in this scenario, {\sys} maintains stable performance.
These experiments demonstrate {\sys}'s ability to tolerate temporal load imbalance and to maintain stable throughput.

\subsection{Directory Aggregation Overhead}
\label{eval:Aggregation Performance}

\begin{figure}[t]
    \centering
    \small
    \begin{subfigure}[b]{0.495\linewidth}
        \centering
        \includegraphics[width=\linewidth]{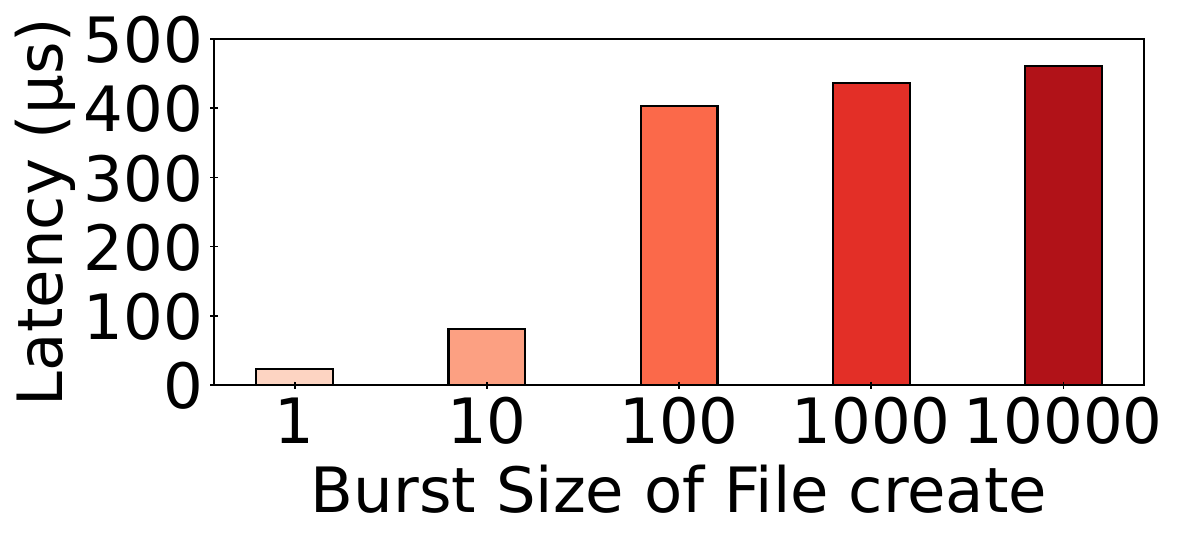}
        \vspace{-15pt}
        \caption{Latency w.r.t burst size on eight servers.}
        \label{fig:fix server}
    \end{subfigure}
    \begin{subfigure}[b]{0.495\linewidth}
        \centering
        \includegraphics[width=\linewidth]{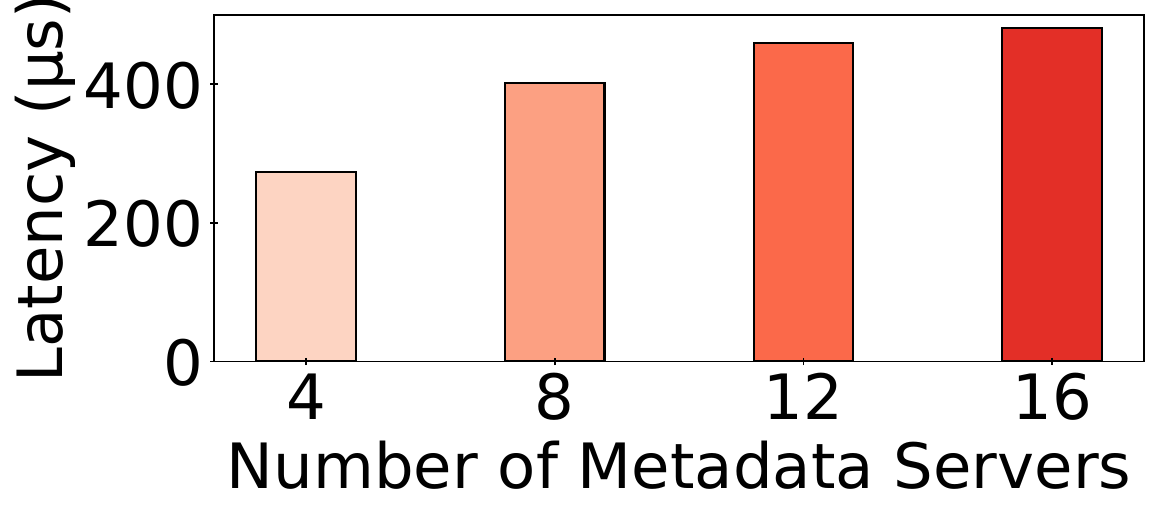}
        \vspace{-15pt}
        \caption{Latency w.r.t. server number with 100 preceding \emph{create}s.}
        \label{fig:fix ratio}
    \end{subfigure}
    \caption{\textbf{Latency of \emph{statdir} after successive \emph{create}.}}
    \label{fig:aggregation}
\end{figure}

Reading a scattered directory in {\sys} requires metadata aggregation.
To evaluate its overhead, we measure the latency of the \emph{statdir} operation after a sequence of file \emph{create} in the directory.
During the \emph{statdir} operation, change-log entries generated by previous \emph{create} operations are aggregated and applied to the directory.
We vary both the number of servers and the number of preceding \emph{create} operations.

\autoref{fig:fix server} shows how the latency of \emph{statdir} increases with a growing number of preceding \emph{create}s, eventually converging at approximately 500\,$\mu$s.
The reason is that as the number of preceding \emph{create}s increases, more change-log entries must be processed during aggregation.
However, the number of entries per server remains bounded (to 29 in our implementation) since the servers proactively push local change-log entries to the inode once they can fill an MTU.

\autoref{fig:fix ratio} demonstrates that the average \emph{statdir} latency increases with the number of servers.
As it increases, more entries can be kept in the change-logs before proactive aggregation is triggered, resulting in more entries being aggregated during \emph{statdir}.

\subsection{End-to-End Performance}
\label{eval:realworld}

We evaluate the end-to-end performance under real-world workloads, with data access included, using a synthetic workload based on realistic operation ratios and two real-world traces.
This experiment uses eight metadata servers and eight data nodes, and the client issues 256 in-flight requests.
No dirty-set overflow occurs in this evaluation.
\autoref{tab:ratio} summarizes the operation ratios for each workload.

The synthetic workload is derived from the operation ratios of traces in Alibaba's deployed PanguFS, which serves various data center services including data processing, object storage, and block storage~\cite{Lv2022InfiniFS}.
Based on these ratios, we perform 10 million operations in 1024 directories, with 80\% of the operations in 20\% of the directories to simulate workload skew.
Data access is omitted from the synthetic workload because the corresponding operation ratio is unavailable.

The CV Training and Thumbnail workloads are representative many-small-file workloads. %
The CV Training workload is traced from training ALEXNET model~\cite{NIPS2012-c399862d,ALEXNET} on the ImageNet dataset~\cite{imagenet-dataset}, which contains 1.28 million files (mostly under 256KB) grouped into 1000 directories.
This trace captures the entire lifecycle of the dataset, including download, access, and removal.
The Thumbnail workload records the process of accessing 1 million images (mostly under 256KB) and creating thumbnails.
We replay the trace with data access enabled and disabled.

\begin{figure}[tb]
    \centering
    \includegraphics[width=\linewidth]{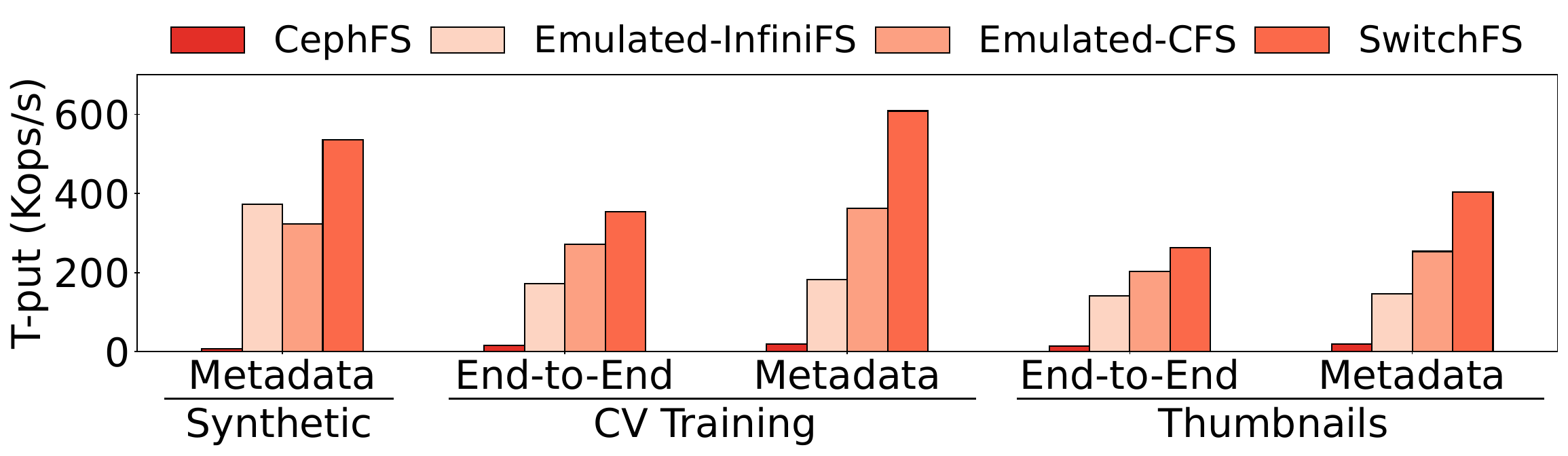}
    \vspace{-15pt}
    \caption{\textbf{Throughput of end-to-end workloads.}}
    \vspace{-5pt}
    \label{fig:realworld}
\end{figure}

As \autoref{fig:realworld} shows,
{\sys}'s improvement in metadata throughput translates into considerable gains in end-to-end throughput.
Compared to CephFS, E-InfiniFS, and E-CFS, {\sys} delivers speedups of up to 76.3$\times$, 2.1$\times$, and 0.7$\times$ for metadata throughput, and up to 21.1$\times$, 1.1$\times$, and 0.3$\times$ for end-to-end performance.

\begin{table}[tb]
    \centering
    \footnotesize
    \resizebox{\linewidth}{!}{%
    \begin{tabular}{ll}
        \hline
        \textbf{Workload}  & \textbf{Operation Ratio} \\ \hline
        \begin{tabular}[c]{@{}l@{}}Data Center\\ Services~\cite{Lv2022InfiniFS}\end{tabular} & \begin{tabular}[c]{@{}l@{}}52.6\% open/close, 12.4\% stat, 9.58\% create,  11.9\% delete, \\9.3\% file rename, 0.1\% chmod, 3.9\% readdir, 0.2\% statdir\end{tabular} \\ \hline
        \begin{tabular}[c]{@{}l@{}}CNN Training\end{tabular}  & \begin{tabular}[c]{@{}l@{}}42.8\% open/close, 21.4\% stat, 14.2\% read, 7.1\% write, 7.1\% create,\\ 7.1\% delete, 0.1\% mkdir, 0.1\% rmdir, 0.1\% statdir, 0.1\% readdir\end{tabular}  \\ \hline
        \begin{tabular}[c]{@{}l@{}}Thumbnail\end{tabular}  & \begin{tabular}[c]{@{}l@{}} 43.9\% open/close, 21.9\% stat, 12.2\% read, 10.9\% write, 10.9\% create,\\ 0.1\% mkdir, 0.1\% statdir, 0.1\% readdir\end{tabular}  \\ \hline
    \end{tabular}
      }
    \caption{\textbf{Metadata operation ratios in real-world traces.}}
    \label{tab:ratio}
\end{table}
    
\subsection{Crash Recovery Time}
We evaluate {\sys}'s recovery time after a server failure and a switch failure.
We simulate each type of failure after creating 10 million files in 100 thousand directories.
{\sys} uses eight servers.
After a server failure, it takes the crashed server 5.77 seconds to recover $\sim$1.25 million inodes to its key-value store and $\sim$1.25 million not-yet-applied change-log entries.
After a switch failure, it takes 3.82 seconds for all the servers to flush not-yet-applied change-log entries to restore {\sys} into a consistent state after the switch reboots.
Note that the recovery time is proportional to the number of operations to be recovered, and could be substantially reduced through the use of checkpointing.
During the recovery process, the end-to-end throughput is zero.

\section{Related Work}%
\label{sec:related}

\paragraph{Metadata partitioning.}
Early DFSs such as GFS~\cite{Ghemawat2003The-Google}, HDFS~\cite{Shvachko2010The-Hadoop}, Farsite~\cite{Farsite}, and QFS~\cite{QFS} manage all metadata on a single metadata server, resulting in poor scalability.
More recent DFSs partition the namespace across multiple metadata servers to enable scalable metadata management.
Some DFSs partition the directory tree at subtree granularity, such as AFS~\cite{10.1145/37499.37500} and CephFS~\cite{Weil2006Ceph}.
While subtree partitioning preserves metadata locality, it is prone to load imbalance~\cite{7832795,Wang2021Lunule}.
Some other DFSs use per-directory partitioning, such as BeeGFS~\cite{BeeGFS}, IndexFS~\cite{Ren2014IndexFS,GIGA,TableFS}, HopsFS~\cite{Niazi2017HopsFS}, Tectonic~\cite{Pan2021Facebooktextquoterights}, and InfiniFS~\cite{Lv2022InfiniFS}.
Since they still group files' inodes with parent directories, they also suffer load imbalance when operations are intensively performed in a few directories.
CFS~\cite{CFS} adopts per-file partitioning, which achieves good load balance but incurs cross-server coordination overhead for double-inode operations.

Previous DFSs have to trade off between load balance and operation overhead when choosing partitioning granularity.
In contrast, {\sys} achieves good load balance by partitioning at per-file granularity while hiding cross-server coordination overhead through asynchronous metadata updates.
In this way, {\sys} achieves the best of both worlds.

\paragraph{Directory contention mitigation.}
Prior DFSs explored how to mitigate directory contention.
LocoFS~\cite{Li2017LocoFS} relaxes POSIX semantics to avoid updating directory metadata for double-inode operations.
DeltaFS~\cite{DeltaFS} does not provide a globally shared namespace but asks clients to explicitly manage synchronization scope, reducing the contention domain.
CFS~\cite{CFS} requires cross-server coordination for double-inode operations while using ordered updates and database primitives to prune the critical section of directory updates.
SingularFS~\cite{SingularFS} stores metadata on non-volatile main memory (NVMM) on a single server and updates directories with atomic instructions to reduce contention.
{\sys} provides a POSIX-compliant global namespace and does not rely on NVMM.
{\sys}'s change-log compaction merges concurrent directory updates, eliminating the single-point serialization for directory updates, and can be combined with previous techniques.

\paragraph{In-network acceleration.}

There are previous studies that explored in-network coordination in distributed systems,
including key-value cache~\cite{Li2016Be-Fast,Jin2017NetCache,Jin2018NetChain,Liu2017IncBricks,PKache,288778,Seer,Liu2019DistCache,Li2020Pegasus,294799,NetRS}, in-network computation~\cite{10.1145/3575693.3575708,10.1145/3582016.3582037,ATP,A2TP,265065,10.1145/3152434.3152461}, distributed memory~\cite{Lee2021MIND,264822}, distributed locks~\cite{Yu2020NetLock,FISSLOCK}, consensus and concurrency control~\cite{9095258,NetPaxos,Li2017Eris,199299,10.14778/3523210.3523213,Harmonia,FLAIR}, and scheduling~\cite{10.1145/3600006.3613170,Draconis}.
{\sys}'s in-network structure is similar to that of Harmonia~\cite{Harmonia}.
However, Harmonia targets in-network conflict detection for fast key-value store replication, while {\sys} adopts in-network state tracking for zero-cost asynchronous DFS operations.
To our knowledge, {\sys} is the first to introduce in-network optimization to distributed filesystems.

\paragraph{Asynchronous update.}
There are previous studies~\cite{10.1145/1394441.1394442,openat2,DeltaFS} that adopted asynchronous updates for data and metadata, and {\sys} is different from them.
XSYNCFS~\cite{10.1145/1394441.1394442} and POSIX-data-write~\cite{openat2} adopt asynchronous data updates, but their approaches are not applicable for metadata, as they allow update loss after crashes, whereas POSIX semantics forbid any loss of metadata updates. In contrast, {\sys}'s asynchronous protocol is persistent and consistent.
DeltaFS~\cite{DeltaFS} explored asynchronous metadata synchronization between clients.
However, it uses explicit APIs to “get” and “publish” metadata updates, leaving the synchronization burden to users.
In contrast, by introducing careful protocol design and in-network coordination, {\sys}'s asynchronous metadata operations are transparent to users and POSIX-compatible.

\section{Conclusion}

This paper proposes {\sys}, a distributed filesystem with asynchronous metadata update to resolve the trade-off between operation efficiency and load balancing.
{\sys} leverages a programmable switch to track directory states in the network, and
adopts change-log compaction to alleviate contention over directory metadata.
Evaluation shows that {\sys} outperforms state-of-the-art systems.

\section*{Acknowledgements}
We sincerely thank our shepherd, Marc Shapiro, and the anonymous reviewers for their constructive comments and insightful suggestions.
This work is supported in part by the National Key R\&D Program of China (2024YFB4506200), the National Natural Science Foundation of China (No. 62132014), the Fundamental Research Funds for the Central Universities, Fundamental and Interdisciplinary Disciplines Breakthrough Plan of the Ministry of Education of China (JYB2025XDXM\allowbreak113), and Huawei Technologies.
The corresponding author is Mingkai Dong (\url{mingkaidong@sjtu.edu.cn}).

\bibliographystyle{ACM-Reference-Format}
\bibliography{references}

\appendix
\section{Appendix}
\label{sec:appendix}
In the appendix, we provide a detailed discussion of the correctness and consistency of {\sys}.

\subsection{Correctness After a Crash}
\label{sec:appendix-crash}
In this subsection, we discuss how {\sys} maintains consistency after a server crash, a switch crash, or both, and provide a concrete example of handling server failure during aggregation.

\paragraph{Switch failure.}
When a switch recovers from failure, it initializes an empty dirty set, and servers aggregate all directories.
Thereby, {\sys} recovers to a consistent state.

\paragraph{Server failure.}
When a server recovers from failure, it recovers its volatile states as follows: it recovers the key-value store and change-log from its local WAL, and recovers the invalidation by cloning that on the other servers.
Note that each not-yet-aggregated directory update is stored in the WAL of either the change-log's or the directory's owner server, so no updates are lost.
The server also proactively aggregates all directories it owns to ensure any interrupted aggregations it issued before the crash (whose corresponding fingerprint may have been removed on the switch) are executed to completion, so that the on-switch dirty set correctly reflects the states of directories.

\paragraph{Example of server failure during aggregation.}
Let's consider an example of server failures during aggregation, involving Server-A (which hosts directory D's inode) and Server-B (which hosts one of D's change-logs).

Consider that at the time one or both servers crash, some change-log entries may have been written to Server-A's WAL, and a subset of those are marked as ``applied'' in Server-B's WAL.

If Server-A crashes, it applies the change-log entries that have been written into Server-A's WAL to D's inode.
Then, Server A proactively aggregates all directories it owns, including D, retrieving the remaining change-log entries of D from Server-B.

If Server-B crashes, Server-A's aggregation times out (due to not receiving a signal from Server-B that all change-log entries have been retrieved) and begins retrying.
Server-B then recovers all change-log entries not marked as “applied” from its WAL, and Server-A aggregates them upon retrying.
Notably, Server-A may receive duplicate entries and leverages entry IDs to ensure each entry is processed only once.

If both servers crash, Server-A applies the change-log entries that have been written into Server-A's WAL to D's inode, and Server-B recovers the remaining change-log entries from Server-B's WAL.
Then, Server A proactively aggregates all directories it owns, including D, retrieving all the change-log entries of D.

\paragraph{Idempotence of recovery.}
The recovery procedures described above are idempotent.
First, the recovery of volatile states (i.e., rebuilding the key-value store, the change-log, and the invalidation list) is idempotent, as it always starts from a clean state after a crash.
Second, the aggregation is idempotent, because:
(a) A directory owner server persists a change-log entry in its WAL before the sender server marks it as ``applied'' in its WAL, ensuring that no change-log entry is lost.
(b) Repeated aggregation of the same change-log entry does not compromise correctness, as servers check entry IDs and ensure each entry is processed only once.
Therefore, nested crashes do not compromise correctness.

\subsection{Operation Consistency and Properties}
\label{sec:appendix-correctness}

In this subsection, we first demonstrate that {\sys} is serializable and preserves the real-time order in a crash-free environment.
Then, we show that {\sys} maintains serializability in the presence of crashes.

\subsubsection{Consistency in a Crash-Free Environment}
We demonstrate that {\sys} ensures the following two properties, when no crash occurs:
\begin{itemize}
    \item Metadata operations are serializable.
    \item Any metadata operation $\beta$ that is issued after operation $\alpha$ returns can see the changes made by $\alpha$.
\end{itemize}

\paragraph{Property 1:} \emph{Metadata operations are serializable.}

In {\sys}, synchronous operations (e.g., open, read, stat) are serialized through locking as in a traditional DFS, and two asynchronous updates (e.g., \emph{mkdir}, \emph{create}) are serialized according to the order in which they obtain the write lock on the parent directory's change-log.
We discuss how asynchronous updates and directory reads are serialized below.

Consider two operations, $\alpha$ and $\beta$: $\alpha$ performs an asynchronous update to a directory, and $\beta$ reads the same directory.
Without loss of generality, let $\alpha$ be \emph{create(/a/b)} and $\beta$ be \emph{readdir(/a)}.
Assume that Server-A hosts the inode of directory \emph{/a}, and Server-B hosts the inode of file \emph{/a/b}.

The executions of $\alpha$ and $\beta$ may interleave in several ways.
We classify these interleavings by whether $\beta$ observes directory \emph{/a} as \emph{scattered} or \emph{normal} in the on-switch dirty set.

\begin{figure}[t]
    \centering
    \begin{subfigure}{\linewidth}
        \centering
        \includegraphics[width=\linewidth]{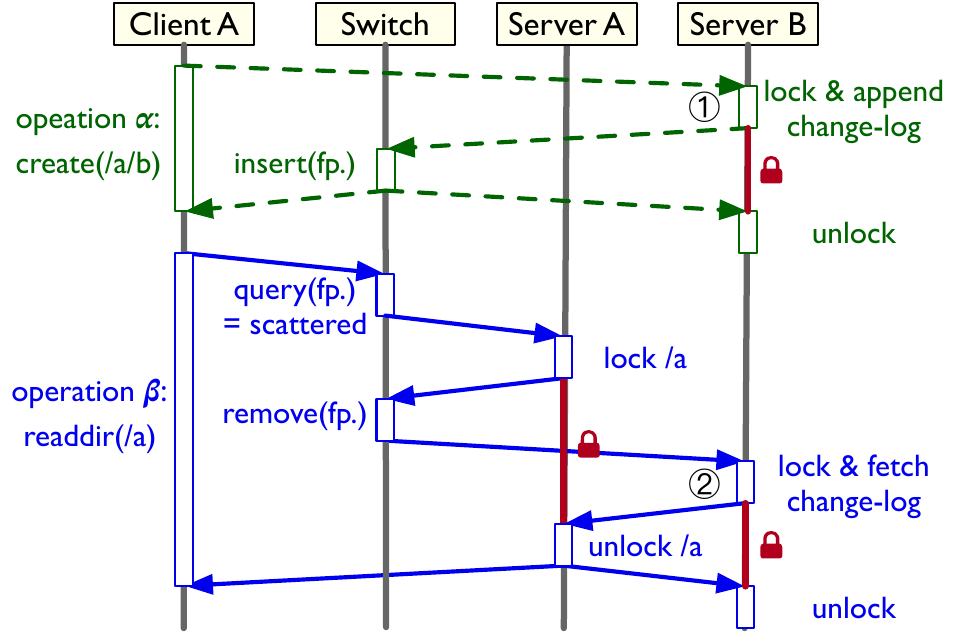}
        \caption{\emph{readdir(/a)} is serialized after \emph{create(/a/b)}.}
        \label{fig:appendix-1-a}
        \vspace{10pt}
    \end{subfigure}
    \begin{subfigure}{\linewidth}
        \centering
        \includegraphics[width=\linewidth]{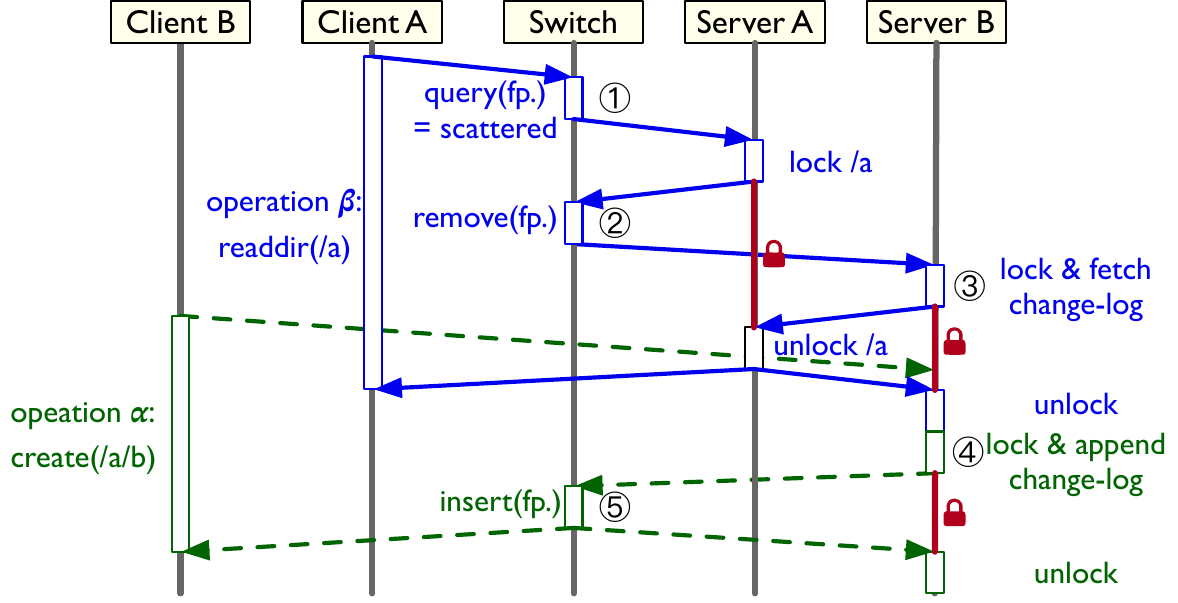}
        \caption{\emph{readdir(/a)} is serialized before \emph{create(/a/b)}.}
        \label{fig:appendix-1-b}
    \end{subfigure}
    \caption{\textbf{Two possible interleavings between \emph{create(/a/b)} and \emph{readdir(/a)} when the dirty set query returns \emph{scattered}.} Numbers in cycles indicate the causal order of events.}
    \label{fig:appendix-1}
\end{figure}

\paragraph{Case 1: $\beta$ finds \emph{/a} marked as \emph{scattered} during its query of the dirty set.}
Depending on whether $\alpha$ updates the change-log on Server-B before or after $\beta$ acquires the lock, two sub-cases arise.
\autoref{fig:appendix-1} illustrates their respective workflows.

\begin{itemize}
    \item \textbf{Case 1.a}: $\alpha$ updates the change-log (\ding{192}) before $\beta$ acquires the lock (\ding{194}), as shown in \autoref{fig:appendix-1-a}.
    In this case, $\beta$ fetches $\alpha$'s update during aggregation and thus observes its effect, serializing $\alpha$ before $\beta$.

    \item \textbf{Case 1.b}: $\alpha$ updates the change-log (\ding{195}) after $\beta$ acquires the lock (\ding{194}), as shown in \autoref{fig:appendix-1-b}.
    In this case, $\beta$ does not fetch $\alpha$'s update during aggregation and therefore does not observe its effect, serializing $\alpha$ after $\beta$.
    Notably, a chain of happen-before relations exists in this interleaving:
    (a) $\alpha$ inserts the fingerprint of \emph{/a} into the dirty set (\ding{196}) after appending the change-log (\ding{195}), and (b) $\beta$ removes the fingerprint from the dirty set (\ding{193}) before acquiring the change-log lock (\ding{194}).
    Therefore, $\alpha$'s insertion of the fingerprint (\ding{196}) occurs strictly after $\beta$'s removal of it (\ding{193}).
    As a result, a subsequent directory read will detect the fingerprint inserted by $\alpha$ and initiate an aggregation that retrieves $\alpha$'s update, thereby making $\alpha$'s effect visible.
\end{itemize}

\begin{figure}[t]
    \centering
    \begin{subfigure}{\linewidth}
        \centering
        \includegraphics[width=\linewidth]{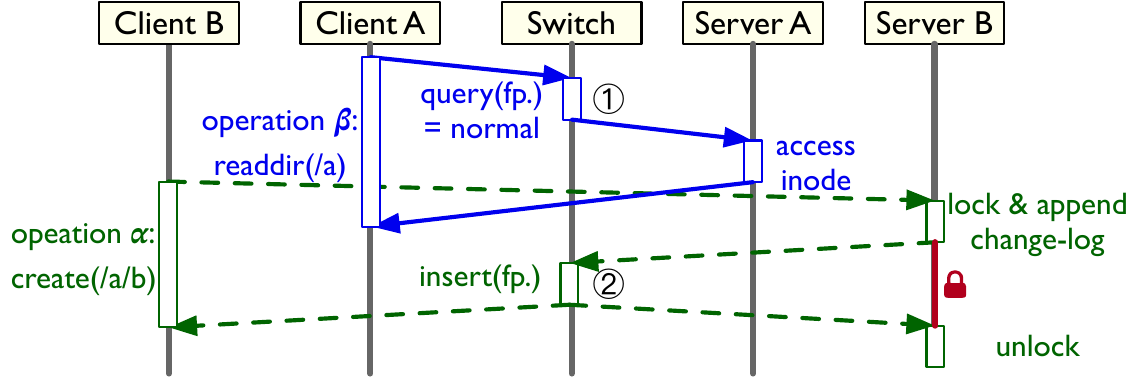}
        \caption{\emph{readdir(/a)} serialized before \emph{create(/a/b)}.}
        \label{fig:appendix-2-a}
        \vspace{10pt}
    \end{subfigure}
    \begin{subfigure}{\linewidth}
        \centering
        \includegraphics[width=\linewidth]{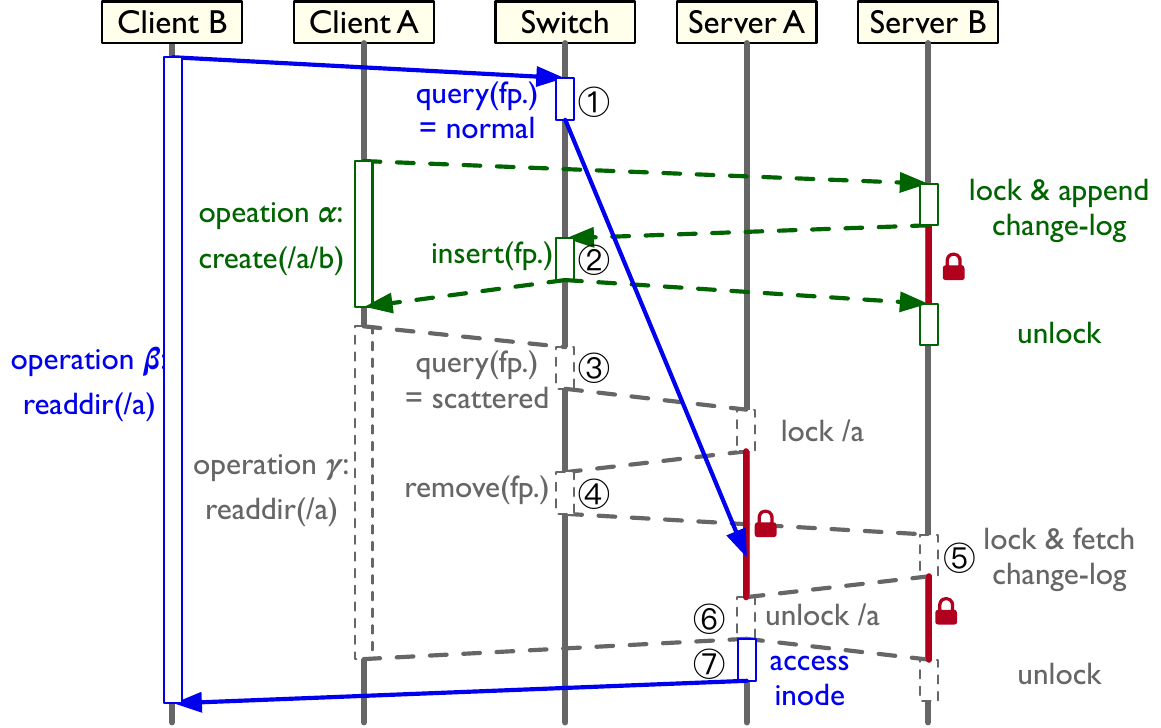}
        \caption{\emph{readdir(/a)} serialized after \emph{create(/a/b)}.}
        \label{fig:appendix-2-b}
    \end{subfigure}
    \caption{\textbf{Two possible interleavings between \emph{create(/a/b)} and \emph{readdir(/a)} when the dirty set query returns \emph{normal}.} Numbers in cycles indicate the causal order of events.}
    \label{fig:appendix-2}
\end{figure}

\begin{figure}[t]
    \centering
    \includegraphics[width=\linewidth]{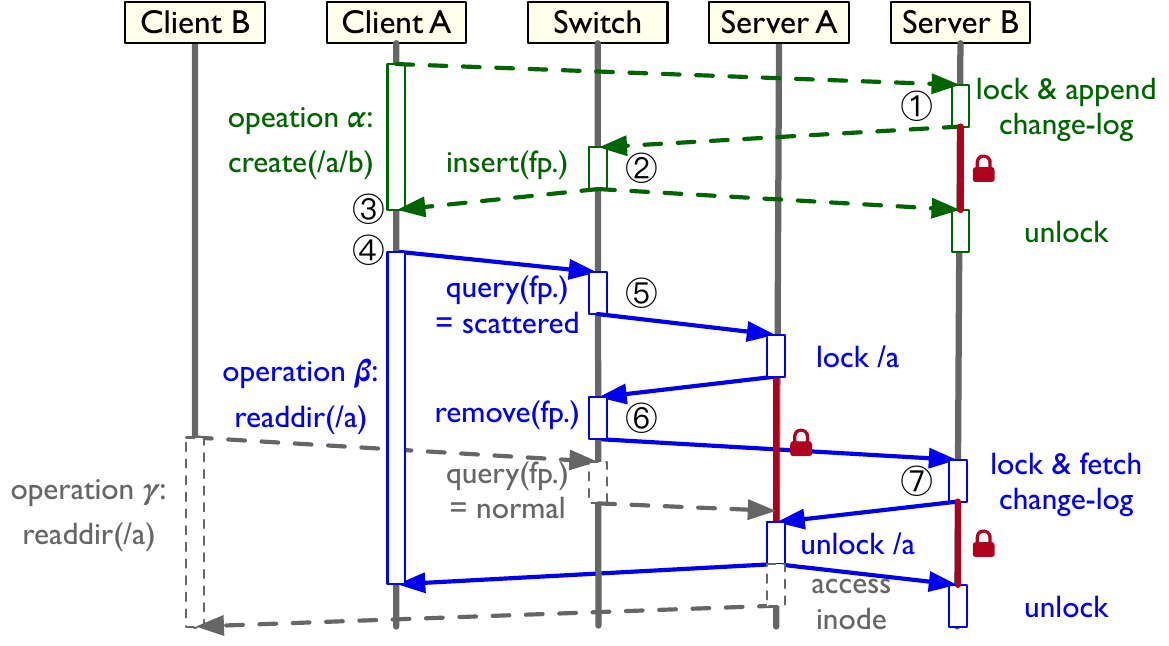}
    \caption{\textbf{Directory reads issued after a directory update can see the effect of the latter.} In this figure, operation $\alpha$ is a directory update, and operations $\beta$, $\gamma$ are directory reads. Numbers in cycles indicate the causal order of events.}
    \label{fig:appendix-3}
\end{figure}

\paragraph{Case 2: $\beta$ finds \emph{/a} marked as \emph{normal} during its query of the dirty set.}
Depending on whether $\beta$ observes the effect of $\alpha$, two sub-cases arise.
\autoref{fig:appendix-2} illustrates their respective workflows.

\begin{itemize}
    \item \textbf{Case 2.a:} $\beta$ reads the inode of directory \emph{/a} on Server-A and returns without observing $\alpha$'s update, as shown in \autoref{fig:appendix-2-a}. In this case, $\alpha$ is serialized after $\beta$.
    Notably, since $\beta$ finds \emph{/a} marked as "normal" in the dirty set, $\beta$ must have queried the dirty set (\ding{192}) before $\alpha$ inserts the \emph{fingerprint} of \emph{/a} (\ding{193}).

    \item \textbf{Case 2.b:} Another operation $\gamma$ triggers an aggregation after $\alpha$ updates the change-log but before $\beta$ reads the inode, as shown in \autoref{fig:appendix-2-b}. Consequently, $\beta$ observes the effect of $\alpha$, and $\alpha$ is serialized before $\beta$.
    Specifically, $\beta$ queries the dirty set (\ding{192}) before $\alpha$ inserts the \emph{fingerprint} of \emph{/a} (\ding{193}). Subsequently, $\gamma$ queries the dirty set (\ding{194}) and initiates an aggregation on directory \emph{/a} (\ding{195}), during which the inode of \emph{/a} is locked.
    $\beta$'s request arrives at Server-A after $\gamma$ acquires the lock, waits for the aggregation to complete (\ding{196}, \ding{197}, \ding{198}), then reads the inode of \emph{/a} and returns (\ding{199}).
    This case is possible due to the potential arbitrary reordering of packets in the network.
\end{itemize}

In all the scenarios discussed above, operations are correctly serialized, eliminating any Time-Of-Check-To-Time-Of-Use (TOCTTOU) issues.

\paragraph{Property 2:} \emph{Any metadata operation $\beta$ that is issued after operation $\alpha$ returns can see the changes made by $\alpha$.}

In {\sys}, since synchronous operations apply in-place updates to objects before returning, their effects are visible to subsequent operations, as in traditional DFSs.
Below, we explain how asynchronous updates and directory reads preserve this property.

In \autoref{fig:appendix-3}, operation $\alpha$ performs an asynchronous directory update on directory \emph{/a}, while operations $\beta$ and $\gamma$ read directory \emph{/a}.
We classify directory reads issued after $\alpha$ returns into two categories based on whether they observe the directory as \emph{scattered} or \emph{normal} in the dirty set, and operations $\beta$ and $\gamma$ are examples.
We illustrate why both categories must observe the changes made by $\alpha$ below.

\begin{itemize}
    \item If the directory read (i.e., operation $\beta$) finds the directory marked as \emph{scattered} during its query of the dirty set (\ding{196}), it triggers an aggregation that fetches updates for \emph{/a}.
    According to the causal order of events marked in \autoref{fig:appendix-3}, $\beta$ must acquire the lock on the change-log (\ding{198}) after $\alpha$ updates it (\ding{192}).
    Therefore, as discussed in Case 1.a of Property 1, $\beta$ is guaranteed to see the effect of $\alpha$.

    \item If the directory read (i.e., operation $\gamma$) finds the directory marked as \emph{normal} during its query of the dirty set, it indicates that the \emph{fingerprint} of the directory has been removed from the dirty set by an earlier aggregation (\ding{197}).
    Since the aggregation blocks directory reads until it completes, $\gamma$ will see the directory updates fetched by the aggregation, including $\alpha$'s update.
\end{itemize}

\subsubsection{Consistency in the Presence of Crashes}
We demonstrate that operations are serializable in the presence of crashes, as follows.

When a crash occurs, interrupted operations do not return, and clients retry them after recovery.
Interrupted metadata reads and uncommitted updates are invisible to clients and leave no side effects, thus not affecting operation serializability. 
We only need to consider the order of committed metadata update operations after recovery.

The recovery procedures in \autoref{sec:appendix-crash} recover operations in the same order as they were committed (written into the WAL) before the crash.
Since operations acquire locks before committing (\autoref{sec:design-asynchronous-metadata-operations}), the commit order is serialized by locks.
Given that locks also serialize the order of metadata-write operations in a crash-free environment (see discussions under property 1 in \autoref{sec:appendix-correctness}), the operation order with and without a crash is consistent.
Therefore, operations are serializable in the presence of crashes.

\subsection{Fault Tolerance of Cluster Reconfiguration}
\label{sec:appendix-reconfiguration}
The procedure of cluster reconfiguration in \autoref{sec:discussion} is fault-tolerant.
We first present the detailed reconfiguration procedure, then discuss its fault tolerance.

\paragraph{Reconfiguration procedure.}
{\sys} employs a centralized coordinator to manage cluster configuration changes.
When servers are added or removed, the coordinator carries out the reconfiguration in three steps:

\begin{enumerate}
\item The coordinator notifies all servers to stop serving normal requests and to aggregate the directories they currently own.

\item Once all servers acknowledge the completion of the aggregation, the coordinator notifies them of the new configuration.
Each server then determines the set of inodes that it no longer owns according to the new configuration and consistent hashing, and migrate these inodes to the new owners via two-phase commit.

\item After all servers acknowledge completing the migration, the coordinator notifies them to resume serving requests.

\end{enumerate}

\paragraph{Fault handling.}
Each step of the reconfiguration is designed to be idempotent.
The coordinator persistently records the start and completion of each step in its local WAL.
If the coordinator crashes during reconfiguration, it asks all the servers to re-execute the reconfiguration from the last step that was not complete.
If a server crashes, it first restores its volatile state by replaying its WAL, then obtains the current reconfiguration step from the coordinator, and re-executes that step.

\end{document}